\PassOptionsToPackage{table}{xcolor}

\documentclass[acmsmall,nonacm]{acmart}

\acmJournal{TAAS}
\acmVolume{0}
\acmNumber{0}
\acmArticle{0}
\acmYear{2026}
\acmMonth{0}
\copyrightyear{2026}
\setcopyright{none}                   
\usepackage{amsmath, amssymb}
\usepackage{graphicx}
\usepackage{algorithm}
\usepackage{algpseudocode}
\usepackage{url}
\usepackage{xcolor}
\usepackage{booktabs}
\usepackage{tabularx}
\usepackage{adjustbox}
\usepackage[capitalize]{cleveref}
\crefname{lstlisting}{Listing}{Listings}
\Crefname{lstlisting}{Listing}{Listings}
\usepackage{subcaption}
\usepackage{multirow}
\usepackage{listings}
\lstdefinestyle{prompt}{
  basicstyle=\small\ttfamily,
  breaklines=true,
  breakatwhitespace=false,
  backgroundcolor=\color{yellow!8},
  frame=single,
  rulecolor=\color{black!40},
  xleftmargin=4pt,
  xrightmargin=4pt,
  aboveskip=6pt,
  belowskip=4pt,
  columns=fullflexible,
  keepspaces=true,
}
\usepackage{tikz}
\usetikzlibrary{positioning, shapes.geometric, calc}

\newcommand{\selfcite}[1]{\cite{#1}}
\newcommand{\selfurl}{\url{https://github.com/lloven/neural-router-experiments}}

\title{Neural Router: Semantic Content Matching for Agentic AI}

\author{Lauri Lov\'en}
\orcid{0000-0001-9475-4839}
\email{lauri.loven@oulu.fi}
\affiliation{%
  \institution{Future Computing Group, University of Oulu}
  \city{Oulu}
  \country{Finland}
}

\author{Abhishek Kumar}
\affiliation{%
  \institution{University of Jyv\"askyl\"a}
  \city{Jyv\"askyl\"a}
  \country{Finland}
}

\author{Alexander Engelhardt}
\affiliation{%
  \institution{Department of Computer Science, University of Helsinki}
  \city{Helsinki}
  \country{Finland}
}

\author{Alaa Saleh}
\affiliation{%
  \institution{Department of Computer Science, University of Helsinki}
  \city{Helsinki}
  \country{Finland}
}

\author{Roberto Morabito}
\affiliation{%
  \institution{Department of Communication Systems, EURECOM}
  \city{Biot}
  \country{France}
}

\author{Xiaoli Liu}
\affiliation{%
  \institution{Department of Computer Science, University of Helsinki}
  \city{Helsinki}
  \country{Finland}
}

\author{Naser Hossein Motlagh}
\affiliation{%
  \institution{Department of Computer Science, University of Helsinki}
  \city{Helsinki}
  \country{Finland}
}

\author{Sasu Tarkoma}
\affiliation{%
  \institution{Future Computing Group, University of Oulu}
  \city{Oulu}
  \country{Finland}
}
\affiliation{%
  \institution{Department of Computer Science, University of Helsinki}
  \city{Helsinki}
  \country{Finland}
}

\renewcommand{\shortauthors}{Lov\'en et al.}

\begin{abstract}
Large language models (LLMs) can serve as the semantic-matching engine of a content-based publish/subscribe broker for agentic AI across the edge--cloud computing continuum, bridging the vocabulary and modality gaps that defeat keyword and embedding filters. Framed as offline multi-label retrieval over three public datasets spanning social-media, legal, and smart-home sensor domains (six LLMs, seven baselines), our central contribution is a two-crossover cost--accuracy characterisation: an analytical \emph{context-window crossover} below which a \textsc{CoverAndMerge} compression pipeline reduces LLM invocations, and an empirical \emph{discrimination-capacity crossover} above which matching accuracy collapses independently of context budget, by a model-dependent factor of parameter count and training generation. Two findings carry practical weight: above the discrimination crossover, compression cannot recover accuracy and only frontier-scale models clear large subscription sets; and there backend choice dominates configuration choice, so model selection, not pipeline tuning, is the primary operator lever. We accompany this with three composable algorithms and a per-cluster Quality-of-Experience framework for autonomic LLM-tier selection.
\end{abstract}



\begin{document}
\maketitle


\section{Introduction}\label{sec:intro}

As autonomous AI agents increasingly operate across the computing continuum, from cloud services through network-edge infrastructure to IoT devices, they must discover and consume information produced by other agents and services whose data formats, vocabularies, and modalities they do not control~\cite{saleh2025towards,tarkoma2023ainative}; \selfcite{loven2026service}. A monitoring agent may need regulatory alerts written in legal jargon; a smart-environment agent must interpret raw sensor activations as human activities; a social-analytics agent consumes user-generated text in constantly evolving slang. Content-based publish/subscribe (pub/sub) systems~\cite{10.1145/857076.857078,CarzanigaWolf:sigcomm03} provide a natural communication substrate for such loosely coupled, data-driven interactions, but traditional designs rely on keyword matching or attribute-based filters~\cite{CRW:TOCS01, Tarkoma2012} that are increasingly limited as the semantic gap between publishers and subscribers grows; embedding-based semantic-matching pipelines (e.g., Sentence-BERT cosine over the same description vocabulary) provide a natural intermediate, but stop short of the abductive reasoning required when modalities or vocabularies do not align.

This paper studies the matching engine for such a broker as an offline multi-label retrieval problem; deployment-side concerns (online operation, fault tolerance, multi-tenant load) are scoped explicitly in \cref{ssec:limitations}.

Recent advances in Large Language Models (LLMs), built on deep neural architectures such as Transformers~\cite{10.5555/3295222.3295349, DBLP:conf/naacl/DevlinCLT19}, have demonstrated remarkable semantic understanding of text. A growing body of work uses LLMs for \emph{routing}, but the term is overloaded: in the LLM literature, ``routing'' typically means selecting which model should handle a given query~\cite{ong2024routellm, hu2024routerbench, vllmsr2025}, or classifying user intent to trigger a predefined action~\cite{aurelio2024semantic}. These systems route queries \emph{to} models. In contrast, content-based pub/sub requires matching published content \emph{against} subscriber interests, a fundamentally different problem in which the LLM must evaluate whether a piece of content satisfies a natural-language interest description.

To date, no system places an LLM inside the pub/sub matching loop. Surveys on message brokers for generative AI~\cite{saleh2025towards} and pub/sub for edge intelligence~\cite{saleh4872730publish} identify the need to adapt brokers for AI workloads but do not propose LLM-based matching. Recent poster work explores LLM-generated notifications~\cite{isahagian2023publish}, and the AI Interconnect concept envisions LLMs as orchestrators within pub/sub~\cite{tarkoma2023ainative}, but neither replaces the matching function itself.

\begin{figure}[t]
    \centering
    \includegraphics[width=0.7\linewidth]{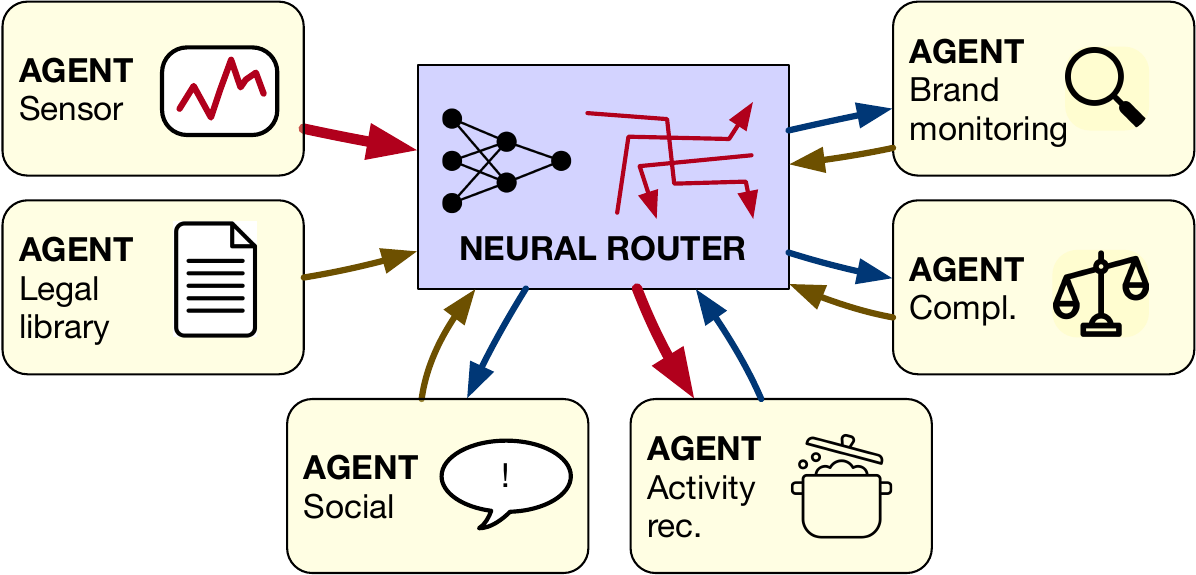}
    \caption{Conceptual overview of the Neural Router. Two pure-publisher
    agents (Sensor, Legal library) and four prosumer agents (Social, Activity
    recognition, Compliance, Brand monitoring) communicate through a single
    broker that matches published content to each agent's natural-language
    interest via LLM reasoning; prosumers both publish and subscribe
    (content/interest in, matched content out). The highlighted link (bold red)
    marks the widest vocabulary/modality gap (raw sensor activations vs.\ a
    plain-language interest) where keyword and embedding matching
    fail by construction. Illustrative, not a deployment; the content types
    correspond to the evaluated domains (\cref{sec:exp}).}
    \label{fig:overview}
\end{figure}

In this paper, we propose the \textbf{Neural Router}~(\cref{fig:overview}), a content-based pub/sub matching engine that makes LLMs a first-class component of the content matching loop.

The design explicitly trades per-invocation latency for semantic accuracy: embedding-based clustering, subscription compression, batching, and parallel LLM invocation collectively keep end-to-end matching latency within the bounds acceptable for analytics, notification, and content-dissemination workloads.
Our contributions are:

\begin{enumerate}
    \item A \emph{cost-accuracy trade-off characterization} for LLM-based content matching with two crossover points: an \emph{analytically derived context-window crossover} predicting when subscription compression becomes cost-effective, and an \emph{empirically identified discrimination-capacity crossover} above which LLM matching accuracy declines, with the magnitude of decline determined by the model's effective capability rather than by parameter count alone (\cref{sec:design}).
    \item A \emph{semantic matching architecture} with three composable algorithms (\textsc{OptimizeSubscriptions}, \\ \textsc{MatchEvents}, \textsc{CoverAndMerge}) and a QoE-based backend assignment mechanism for heterogeneous LLM backends (\cref{sec:design}).
    \item An \emph{empirical validation} across three content domains (CardiffNLP tweets, EUR-Lex legal documents, CASAS smart-home sensor logs), evaluating six LLM backends (open-weight Qwen 2.5 at 1.5B/7B/32B and Mistral 7B; closed-API Claude Haiku and Sonnet) and seven ablation configurations against seven baselines, including zero-shot NLI with DistilBART-MNLI. The results confirm the cost model's polarity: raw matching (A0) dominates \emph{above} the context-window crossover, while below it the compression pipeline's empirical recall is conditional on the merged set's discrimination structure (\cref{ssec:crossover}). The four-model open-weight sweep shows the discrimination-capacity decline at $|\mathcal{S}|{=}201$ is monotonic in parameter count within the Qwen family and is lifted, at fixed scale, by newer training generations (Mistral 7B $>$ Qwen 2.5 7B at matched scale; broader cross-family validation deferred) (\cref{sec:exp,sec:results}).
    \item A \emph{QoE-based heterogeneous backend assignment framework}, with an empirical analysis identifying the \emph{calibration-sample-to-discrimination-gap ratio} as the binding parameter. The framework routes subscription clusters to backends by per-cluster min-max-normalised weighted scoring of accuracy, cost, and latency, with three operator presets (\texttt{accuracy\_first}, \texttt{balanced}, \texttt{cost\_first}) yielding three distinct assignments. On CardiffNLP at a Qwen-2.5 (7B, 32B) gradient in the offline-batch single-tenant regime, round-robin is a strong calibration-free baseline --- closing $83\%$ of the F1 gap at $0.7\times$ the larger-tier latency --- and greedy QoE separates from it only at \texttt{cost\_first}; a calibration-fraction sweep ($\{0.05,\ldots,1.00\}$) never crosses $\alpha{=}0.05$ (Friedman $p{=}0.290$), so the ratio holds only in its weak, cost-side form on this gradient. Value under load skew, throughput heterogeneity, backend failure, and SLA constraints is left to deployment studies (\cref{sec:results,ssec:limitations}).
\end{enumerate}

The Neural Router specifies a single-node content matching engine. The question of how multiple matching engine instances can be federated across a distributed computing continuum (e.g., across 6G network slices) is an important extension that we discuss as future work in \cref{sec:discussion}.

The remainder of this article is organised as follows. \Cref{sec:relwork} surveys related work and positions our contribution. \Cref{sec:design} presents the Neural Router's design, including the problem statement, architecture, algorithms, and cost model. \Cref{sec:exp} describes the experimental setup. \Cref{sec:results} presents the results. \Cref{sec:discussion} discusses limitations, implications, and future directions. \Cref{sec:conclusion} concludes.

\section{Related Work}\label{sec:relwork}

We review five bodies of work that surround the Neural Router: traditional content-based pub/sub (\cref{sec:rw:cbps}), distributed stream processing platforms (\cref{sec:rw:stream}), LLM routing and semantic routing in the machine-learning community (\cref{sec:rw:llmrouting}), recent efforts that bring LLMs closer to publish/subscribe (\cref{sec:rw:llmpubsub}), and autonomic / adaptive systems for the computing continuum (\cref{sec:rw:autonomic}). \Cref{sec:rw:position} positions our contribution.

\subsection{Content-based publish/subscribe}\label{sec:rw:cbps}

Content-based pub/sub matches messages to subscribers by evaluating predicates over message content, rather than relying on pre-assigned topic labels~\cite{10.1145/857076.857078, Jacobsen2009, CRW:TOCS01}. The central challenge is efficient matching: as the number of subscriptions and the dimensionality of the content grow, the broker must evaluate an increasing number of filter predicates per event.

Traditional systems express subscriptions as conjunctions of attribute-value constraints~\cite{CRW:TOCS01, Tarkoma2012}. The algorithmic lineage of efficient subscription matching includes the Yan-Garcia-Molina counting algorithm, SIENA's covering and merging relations~\cite{CarzanigaWolf:sigcomm03}, BE-Tree, JEDI, Hermes, and PADRES~\cite{li2005padres}; \textsc{CoverAndMerge} (\cref{alg:covermerge}) is structurally analogous to SIENA's covering / merging operations, but uses an LLM in place of the syntactic-subsumption test --- a deliberate design choice motivated by the natural-language subscription descriptions targeted by this work. Scalability beyond syntactic-filter regimes has been pursued through space-partitioning techniques that cluster subscriptions hierarchically~\cite{wang2014general} and Kafka extensions with specialised low-latency topic types~\cite{qian2021fat}. Network-level optimisations embed matching into SDN data planes via P4~\cite{wernecke2022evaluating} or OpenFlow~\cite{bhowmik2018expressive}. Privacy-preserving variants apply order-preserving encryption~\cite{li2020privacy} or data-splitting techniques~\cite{denis2020privacy} to enable encrypted matching. Publisher-side filtering has been proposed for edge deployments where bandwidth is constrained~\cite{wang2024shutpub}.

A common limitation of these systems is their reliance on structured, attribute-based filter languages. Subscriptions must be expressed as Boolean predicates over named fields, which limits expressiveness when content is unstructured text, sensor data, or other high-dimensional modalities. Semantic pub/sub systems such as OpenPubSub~\cite{zaarour2022openpubsub} address part of this gap by using embedding-based similarity for routing in peer-to-peer networks, but the matching function remains a fixed cosine threshold rather than a learned or reasoned evaluation.

\subsection{Distributed stream processing}\label{sec:rw:stream}

Modern data infrastructure relies on distributed stream processing platforms that share pub/sub's concern with high-throughput event dissemination but differ in their matching semantics. Apache Kafka~\cite{kreps2011kafka} provides durable, partitioned log-based messaging with topic-based routing and configurable delivery guarantees. Apache Flink~\cite{carbone2015apache} extends the stream processing model with stateful operators, event-time semantics, and exactly-once processing, enabling complex event processing (CEP) over continuous data flows. Ray~\cite{moritz2018ray} provides a general-purpose distributed computing framework that has become a standard substrate for distributed LLM serving and training. Efficient LLM serving systems such as vLLM~\cite{kwon2023vllm} and SGLang~\cite{zheng2024sglang} build on these platforms to optimise memory management and structured generation for high-throughput inference.

These platforms are complementary to the Neural Router: they provide the transport and execution substrate on which a semantic matching broker can be deployed, but they do not themselves perform content-based semantic matching. Kafka routes by topic partition, Flink by user-defined operator logic, and Ray by task scheduling. None evaluates whether an event's \emph{meaning} satisfies a subscriber's \emph{interest}.

\subsection{LLM routing and semantic routing}\label{sec:rw:llmrouting}

In the machine-learning community, ``routing'' refers to selecting which model should handle a given query, optimising cost--quality trade-offs across a pool of LLMs. RouteLLM~\cite{ong2024routellm} trains a router on preference data to decide, per query, whether to invoke a strong or weak model, achieving substantial cost savings on standard benchmarks with minimal quality loss. RouterBench~\cite{hu2024routerbench} provides a systematic benchmark and evaluation framework for such multi-LLM routing systems. The vLLM Semantic Router~\cite{vllmsr2025} extends this idea to serving infrastructure, using semantic embeddings to direct queries to specialised model replicas within the vLLM serving engine.

A related line of work uses semantic similarity for \emph{intent classification}: the Semantic Router framework~\cite{aurelio2024semantic} classifies user utterances into predefined intent categories using embedding distance, triggering deterministic actions without LLM invocation. This is effectively a dispatch mechanism that routes queries to handlers.

All of these systems route queries \emph{to} models or actions. The Neural Router addresses a fundamentally different problem: it uses an LLM to match published content \emph{against} natural-language subscriber interests within a pub/sub broker. The LLM is not the destination of a route but the engine that evaluates the matching function.

\subsection{LLMs in publish/subscribe}\label{sec:rw:llmpubsub}

Several recent works bring LLMs and pub/sub closer together, though none places the LLM inside the matching loop.

Saleh et al.~\cite{saleh2025towards} survey how existing message broker architectures (Kafka, RabbitMQ, MQTT, NATS, Redis) can be adapted to serve generative AI workloads, such as prompt routing, model output distribution, and agentic communication. They identify architectural requirements for AI-native brokers but do not propose LLM-based matching. A companion survey~\cite{saleh4872730publish} reviews pub/sub for edge intelligence, cataloguing how pub/sub has been used in IoT and edge computing and identifying open challenges for AI-driven enhancements.

Isahagian et al.~\cite{isahagian2023publish} demonstrate LLM-enhanced pub/sub in which subscribers express interests in natural language and the system uses an LLM to generate customised notifications from matched publications. Their focus is on notification \emph{generation} (post-match enrichment), not on the matching function itself: the underlying matching still relies on conventional mechanisms.

Tarkoma et al.~\cite{tarkoma2023ainative} propose the AI Interconnect, a multi-layer semantic pub/sub architecture for 6G systems that envisions LLMs as orchestrators for prompt routing, inference result dissemination, and model-update distribution. The LLM orchestrates the pub/sub fabric but does not serve as the content-matching engine.

The Neural Pub/Sub paradigm~\cite{loven2023can} introduces the concept of using neural models for content-based routing in the computing continuum, proposing a distributed architecture in which neural matching replaces traditional predicate evaluation. The present paper instantiates this concept for a single broker, providing the formal problem statement, algorithms, cost model, and experimental evaluation.

\subsection{Autonomic and adaptive systems for the computing continuum}\label{sec:rw:autonomic}

In TAAS terms the Neural Router is an adaptive system: a closed calibration-and-assignment loop (\cref{alg:calibrate}) selects an LLM backend per cluster under operator-preference weights, and the cost-model crossover predicates partition the deployment regime. We position the contribution against the autonomic-computing tradition originating with Kephart and Chess~\cite{kephart2003vision} and the self-adaptive-software taxonomy of Salehie and Tahvildari~\cite{salehie2009self}, which together developed control-theoretic, MAPE-K, requirements-aware, and most recently AI-augmented architectural styles for self-management~\cite{nascimento2023}. Closely related at TAAS, the \emph{MemIndex} framework~\cite{saleh2025memindex} proposes intent-indexed adaptive memory management for LM-based multi-agent pub/sub --- a complementary layer to the broker-level matching function~$\mu$ addressed here. The Neural Router is in the AI-augmented direction (LLM as decision component, per-cluster QoE calibration as adaptation trigger), with the QoE scalarisation echoing QoS-aware service composition~\cite{zeng2003qosaware,ardagna2007adaptive} applied to LLM-tier selection. We do not adopt MAPE-K vocabulary specifically because the calibration loop is one-shot at design time rather than a continuously running cycle; a runtime extension that lifts $W_{\mathrm{cross}}$ and $|\mathcal{S}|_{\mathrm{cross}}$ into online predicates is left to follow-on work. The Edge-HPC/Cloud computing continuum~\cite{beckman_fog,parashar2025everywhere,donta2023exploring,kokkonen2023autonomy} extends these concerns across heterogeneous tiers and argues that autonomic management techniques must evolve to handle the resulting distribution and dynamic behaviour. We engage this framing analytically --- the cost-model crossover and the heterogeneous-backend QoE assignment apply at any tier --- but defer cross-tier deployment to follow-on work; the experimental backends in this paper span HPC (CSC Mahti / Puhti) and cloud (Anthropic API), with the edge-tier $W{=}4$K regime addressed analytically only.

\subsection{Positioning}\label{sec:rw:position}

The five bodies of work above are complementary rather than competing. Traditional content-based pub/sub uses structured filters and provides no semantic understanding; embedding-based variants such as OpenPubSub use a fixed cosine threshold and no reasoning. Stream-processing platforms (Kafka, Flink, Ray) and modern LLM-serving systems (vLLM, SGLang) are transport and execution substrates that do not themselves perform content matching. LLM routing systems (RouteLLM, RouterBench, the vLLM Semantic Router, Aurelio's Semantic Router) select \emph{which model handles a query}, the inverse of our problem. Recent LLM-pub/sub work uses LLMs for orchestration~\cite{tarkoma2023ainative}, post-match notification generation~\cite{isahagian2023publish}, or overall architectural framing~\cite{saleh2025towards, loven2023can}, never as the matching engine itself. The Neural Router occupies this gap: it defines a formal matching function evaluated by an LLM, supported by embedding-based clustering, cover/merge compression, and a context-window-aware cost model for batch processing.

\section{Neural Router Design}\label{sec:design}

\subsection{Problem Statement}\label{ssec:problem}

We consider a content-based publish/subscribe (pub/sub) system in which subscribers express their interests as free-text natural-language descriptions. Let $\mathcal{S} = \{s_1, \ldots, s_n\}$ denote the active subscriptions and $\mathcal{E} = \{e_1, \ldots, e_m\}$ the published events. A \emph{matching function} $\mu(e_j, s_i) \in \{0,1\}$ returns 1 when event~$e_j$ satisfies subscription~$s_i$; the \emph{routing problem} is to compute, for every $e_j$, the set $R(e_j) = \{s_i \mid \mu(e_j, s_i) = 1\}$ and deliver accordingly. The design objective is to maximise F1 subject to latency and cost constraints, encoding an explicit \emph{accuracy--latency trade-off}: LLM-based matching offers richer semantic understanding than syntactic filters, but each invocation carries non-trivial latency (tens to hundreds of milliseconds depending on model and deployment). Traditional pub/sub routers evaluate $\mu$ with Boolean filters or keyword matching, which fail under paraphrase, synonymy, and contextual meaning; the Neural Router addresses this gap by implementing $\mu$ through an LLM, supported by embedding-based clustering to partition the matching workload into tractable units. The targeted regime is one where per-event latency on the order of seconds is acceptable (analytics dashboards, notification feeds, content recommendation, regulatory-alert dissemination); real-time control loops (sub-millisecond matching) are outside scope.

\subsection{Architecture Overview}\label{ssec:arch-overview}

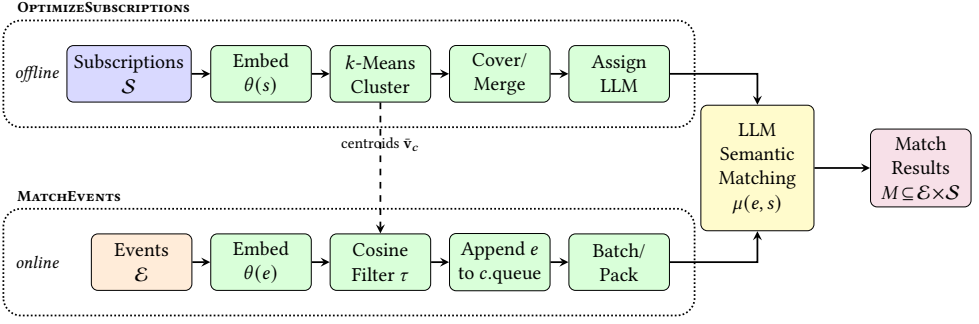
\begin{figure*}[t!]
  \centering
  \resizebox{0.92\textwidth}{!}{%
  \begin{tikzpicture}[
    box/.style={draw, rounded corners=3pt, minimum height=0.9cm, minimum width=1.6cm, align=center, font=\small},
    subbox/.style={box, fill=blue!15},
    evtbox/.style={box, fill=orange!15},
    procbox/.style={box, fill=green!15},
    llmbox/.style={box, fill=yellow!25, minimum height=2.0cm, minimum width=1.8cm},
    outbox/.style={box, fill=purple!12},
    arr/.style={->, >=stealth, thick},
    darr/.style={->, >=stealth, thick, dashed},
    lbl/.style={font=\scriptsize, midway, above},
    lblb/.style={font=\scriptsize, midway, below},
  ]
    \node[subbox] (subs) at (1.8, 1.5) {Subscriptions\\$\mathcal{S}$};
    \node[procbox] (embed_s) at (3.9, 1.5) {Embed\\$\theta(s)$};
    \node[procbox] (cluster) at (5.8, 1.5) {$k$-Means\\Cluster};
    \node[procbox] (covmerge) at (7.7, 1.5) {Cover/\\Merge};
    \node[procbox] (assign) at (9.6, 1.5) {Assign\\LLM};

    \draw[arr] (subs) -- (embed_s);
    \draw[arr] (embed_s) -- (cluster);
    \draw[arr] (cluster) -- (covmerge);
    \draw[arr] (covmerge) -- (assign);

    \node[font=\footnotesize\itshape, anchor=east] at (0.8, 1.5) {offline};

    \node[evtbox] (events) at (2.0, -1.5) {Events\\$\mathcal{E}$};
    \node[procbox] (embed_e) at (3.9, -1.5) {Embed\\$\theta(e)$};
    \node[procbox] (cosine) at (5.8, -1.5) {Cosine\\Filter $\tau$};
    \node[procbox] (append) at (7.7, -1.5) {Append $e$\\to $c.\mathrm{queue}$};
    \node[procbox] (batch) at (9.6, -1.5) {Batch/\\Pack};

    \draw[arr] (events) -- (embed_e);
    \draw[arr] (embed_e) -- (cosine);
    \draw[arr] (cosine) -- (append);
    \draw[arr] (append) -- (batch);

    \node[font=\footnotesize\itshape, anchor=east] at (0.8, -1.5) {online};

    \node[llmbox] (llm) at (11.8, 0) {LLM\\Semantic\\Matching\\$\mu(e,s)$};

    \node[outbox] (output) at (14.4, 0) {Match\\Results\\$M \!\subseteq\! \mathcal{E}\!\times\!\mathcal{S}$};

    \draw[arr] (assign) -| (llm.north);
    \draw[arr] (batch) -| (llm.south);

    \draw[darr] (cluster.south) -- ++(0,-0.9) -| node[lbl, pos=0.25] {centroids $\bar{\mathbf{v}}_c$} (cosine.north);

    \draw[arr] (llm) -- (output);

    \draw[thick, densely dotted, black, rounded corners=8pt] (-0.2, 0.65) rectangle (10.8, 2.35);
    \draw[thick, densely dotted, black, rounded corners=8pt] (-0.2, -2.35) rectangle (10.8, -0.65);

    \node[font=\scriptsize\bfseries, anchor=south west] at (-0.1, 2.35) {\textsc{OptimizeSubscriptions}};
    \node[font=\scriptsize\bfseries, anchor=south west] at (-0.1, -0.65) {\textsc{MatchEvents}};

  \end{tikzpicture}%
  }
  \caption{Neural Router architecture (single-broker view). \emph{Top (offline):} subscriptions are embedded, clustered by semantic similarity, compressed via cover/merge, and assigned to LLM instances (\textsc{OptimizeSubscriptions}). \emph{Bottom (online):} events are embedded and filtered to candidate clusters by cosine similarity to cluster centroids, then batched and packed into LLM prompts (\textsc{MatchEvents}). The LLM evaluates the semantic matching function $\mu(e,s)$ and returns match results.}
  \label{fig:architecture}
\end{figure*}

The Neural Router is a content-based pub/sub broker that replaces filter evaluation with LLM-driven semantic matching (\cref{fig:architecture}). Two processing paths share a common embedding space. The \emph{subscription path} (offline) embeds free-text subscription descriptions with a pre-trained encoder (e.g., Sentence-BERT~\cite{reimers2019sentence}), clusters the embeddings into $k$ semantically coherent groups via $k$-means, compresses each cluster via cover/merge (subscriptions subsumed by a broader one are removed; similar subscriptions are combined), and assigns each cluster to an LLM instance based on QoE criteria such as accuracy and latency budget. The \emph{event path} (online) embeds incoming events, routes them to the nearest cluster(s) by cosine similarity to cluster centroids, accumulates them into batches by time window or count threshold, and packs each batch with the cluster's compressed subscription set into an LLM prompt; the LLM returns event-to-subscription matches.

This two-level architecture (coarse-grained embedding clustering for scale, fine-grained LLM matching for precision) confines each LLM invocation to a semantically coherent subset of the subscription space and keeps prompts within context-window limits.

\textbf{Matching and delivery semantics.}
The matching function $\mu(e,s)$ evaluates whether the semantic content of event~$e$ satisfies the interest expressed in subscription~$s$ rather than relying on topic labels --- the canonical content-based pub/sub semantics, here delegated to an LLM. The current implementation provides \emph{at-most-once} delivery; extending to at-least-once (with retry) or exactly-once (with deduplication) is a transport-layer concern and does not affect the matching algorithms. Because LLM inference is not bitwise deterministic even at temperature~0 (GPU floating-point non-determinism produces occasional variation), the matching function is not guaranteed idempotent across repeated invocations on the same input; we mitigate this with greedy decoding (temperature~0) and structured-output parsing, and deployments needing strict idempotency can cache results keyed on (event hash, subscription-set hash).

\subsection{Algorithms}\label{ssec:algorithms}

The Neural Router's core function is \emph{semantic matching} --- deciding, for each event, which subscriptions it satisfies (forwarding matched content across a network of brokers belongs to the distributed architecture and is outside the scope of this paper). We formalise the matching pipeline through three procedures whose processing flows mirror \cref{fig:architecture}.

\textbf{Subscription Optimisation (\cref{alg:optimize}).}
\textsc{OptimizeSubscriptions} prepares the matching table.
Given subscriptions~$\mathcal{S}$, an embedding model~$\theta$, the number of clusters~$k$, and an LLM pool~$\mathcal{L}$, it
(1)~embeds every subscription,
(2)~partitions the embeddings into $k$ clusters via $k$-means,
(3)~for each cluster selects the LLM that maximises a quality-of-experience (QoE) score (combining accuracy, latency, and cost targets) and invokes LLM-assisted cover/merge compression (\cref{alg:covermerge}), and
(4)~computes the cluster centroid for later event assignment.
The QoE score per cluster $c$ and backend $\ell\in\mathcal{L}_c$ (the set of candidate backends for $c$) is a weighted scalarisation of three per-cluster, per-backend metrics, each min-max normalised across $\mathcal{L}_c$ to neutralise absolute-scale dependence:
\begin{equation}\label{eq:qoe}
  \mathit{QoE}(c, \ell) \;=\; \alpha\,\widetilde{F}_1(c,\ell) \;+\; \beta\,\widetilde{C}(c,\ell) \;+\; \gamma\,\widetilde{L}(c,\ell), \quad \alpha + \beta + \gamma = 1, \;\alpha,\beta,\gamma \ge 0.
\end{equation}
Each component $\widetilde{X}(c,\ell)$ is the min-max-normalised version of its raw counterpart over $\mathcal{L}_c$ --- $\hat{F}_1(c,\ell)$ for accuracy, $p(\ell)\bar{T}(c,\ell)$ for cost, and $\bar{t}_{\mathrm{llm}}(\ell)$ for latency --- with cost and latency inverted (lower-is-better) so all three lie in $[0,1]$ with higher = better, and $0/0 := 0.5$ when a metric is constant across $\mathcal{L}_c$. Three weight presets operationalise operator preference: \texttt{accuracy\_first} $(\alpha,\beta,\gamma)=(0.70,0.15,0.15)$, \texttt{balanced} $(0.34,0.33,0.33)$, \texttt{cost\_first} $(0.15,0.70,0.15)$. This weighted-sum scalarisation places per-cluster backend selection in the QoS-aware service-composition tradition~\cite{zeng2003qosaware,ardagna2007adaptive}, and inherits that tradition's well-known caveat: a linear scalarisation can only reach Pareto-optimal backends on the \emph{convex hull} of the accuracy/cost/latency trade-off surface; a backend that is Pareto-optimal but lies in a non-convex region of the front is unreachable by any weight setting $(\alpha,\beta,\gamma)$. On the two-tier gradient evaluated here the candidate front is effectively convex, so this does not bind; at larger backend pools a Chebyshev or $\varepsilon$-constraint scalarisation would be needed to reach non-convex-front backends, which we leave to follow-on work.

\textbf{Calibration and assignment (\cref{alg:calibrate}).} The per-backend accuracy estimate $\hat{F}_1(c, \ell)$ is obtained through a one-shot calibration phase: each backend processes a 10\,\% held-out sample of cluster $c$'s events, and macro-F1, mean per-event latency $\bar{t}_{\mathrm{llm}}(\ell)$, and expected token count $\bar{T}(c,\ell)$ are recorded. Backend $\ell^*(c) = \arg\max_{\ell \in \mathcal{L}_c} \mathit{QoE}(c,\ell)$ is selected per cluster. Backends with calibrated $\hat{F}_1=0$ across all clusters are filtered before normalisation to avoid trivial argmax pathologies on saturating denominators (this filter activates on the Qwen-1.5B A0 F1${=}0.000$ cell in \cref{ssec:res-cross}). The empirical QoE study under all three presets is reported in \cref{ssec:res-qoe}; analyses of round-robin as a calibration-free baseline and of QoE-optimised under varying calibration-sample fractions are also in that section.

\begin{algorithm}
\caption{Calibrate and assign (QoE).}\label{alg:calibrate}
\begin{algorithmic}[1]
\Procedure{CalibrateAndAssign}{$C, \mathcal{L}, E_\mathrm{cal}, (\alpha,\beta,\gamma)$}
    \For{$c \in C$, $\ell \in \mathcal{L}$}
        \State Run $\ell$ on a 10\% sample $E_\mathrm{cal}^{(c)} \subset E$ assigned to $c$;
        \State \quad record $\hat{F}_1(c,\ell), \bar{t}_{\mathrm{llm}}(\ell), p(\ell)\bar{T}(c,\ell)$.
    \EndFor
    \For{$c \in C$}
        \State $\mathcal{L}_c \gets \{\ell \in \mathcal{L} : \hat{F}_1(c,\ell) > 0\}$ \Comment{Filter saturating denominators}
        \State Compute $\widetilde{F}_1, \widetilde{C}, \widetilde{L}$ over $\mathcal{L}_c$ per Eq.~\ref{eq:qoe}.
        \State $c.\ell \gets \arg\max_{\ell \in \mathcal{L}_c}\; [\alpha \widetilde{F}_1 + \beta \widetilde{C} + \gamma \widetilde{L}]$.
    \EndFor
    \Return $\{c.\ell\}_{c \in C}$
\EndProcedure
\end{algorithmic}
\end{algorithm}

\begin{algorithm}
\caption{Subscription optimisation.}\label{alg:optimize}
\begin{algorithmic}[1]
\Procedure{OptimizeSubscriptions}{$\mathcal{S},\; \theta,\; k,\; \mathcal{L},\; \mathit{reunite}$}
    \State $\mathbf{v} \gets [\,\theta(s) \mid s \in \mathcal{S}\,]$ \Comment{Embed all subscriptions}
    \State $C \gets \Call{KMeans}{\mathbf{v},\; k}$ \Comment{Cluster: $C = \{c_1, \ldots, c_k\}$}
    \For{$c \in C$}
        \State $c.\ell \gets \arg\max_{\ell \in \mathcal{L}}\; \mathit{QoE}(c,\, \ell)$ \Comment{Select LLM for cluster}
        \State $c.\mathcal{S}' \gets \Call{CoverAndMerge}{c.\mathcal{S},\; c.\ell}$ \Comment{Compress subscriptions}
        \State $c.\bar{\mathbf{v}} \gets \text{mean}\{\theta(s) \mid s \in c.\mathcal{S}'\}$ \Comment{Cluster centroid}
    \EndFor
    \If{$\mathit{reunite}$} \Comment{Merge into single search space}
        \State $C \gets \bigl\{\,(\bigcup_{c \in C} c.\mathcal{S}',\; c_1.\ell)\,\bigr\}$
    \EndIf
    \Return $C$
\EndProcedure
\end{algorithmic}
\end{algorithm}

An optional $\mathit{reunite}$ flag (line~8 of \cref{alg:optimize}) merges all compressed clusters back into a single search space after per-cluster compression; its effect is evaluated in the ablation study (\cref{sec:exp}).

\textbf{Event Matching (\cref{alg:match}).}
\textsc{MatchEvents} evaluates the matching function~$\mu$ for a batch of incoming events.
Each event is embedded using the same model~$\theta$, then assigned to every cluster whose centroid similarity exceeds a threshold~$\tau$.
Events accumulate in per-cluster queues; for each non-empty queue, the compressed subscriptions and queued events are packed into an LLM prompt, and the cluster's LLM returns the top-$\kappa$ matching subscriptions per event.

\begin{algorithm}
\caption{Event matching.}\label{alg:match}
\begin{algorithmic}[1]
\Procedure{MatchEvents}{$\mathcal{E},\; C,\; \theta,\; \tau,\; \kappa$}
    \For{$e \in \mathcal{E}$}
        \State $\mathbf{v}_e \gets \theta(e)$ \Comment{Embed event}
        \For{$c \in C$}
            \If{$\cos(\mathbf{v}_e,\; c.\bar{\mathbf{v}}) \geq \tau$} \Comment{Centroid similarity}
                \State $c.\mathit{queue}.\text{append}(e)$
            \EndIf
        \EndFor
    \EndFor
    \State $M \gets \emptyset$ \Comment{Set of matching decisions}
    \For{$c \in C$ \textbf{where} $c.\mathit{queue} \neq \emptyset$}
        \State $p \gets \Call{Pack}{c.\mathcal{S}',\; c.\mathit{queue},\; \kappa}$ \Comment{Build LLM prompt}
        \State $M \gets M \cup \Call{InvokeLLM}{c.\ell,\; p}$ \Comment{$(e, s)$ pairs}
    \EndFor
    \Return $M$
\EndProcedure
\end{algorithmic}
\end{algorithm}

\textbf{LLM-Assisted Cover and Merge (\cref{alg:covermerge}).}
\textsc{Cover-AndMerge} compresses the subscription set of a single cluster using two operations.
\emph{Covering} removes a subscription~$s_j$ whose semantics are entirely subsumed by a broader subscription~$s_i$;
\emph{merging} replaces two overlapping subscriptions with a single combined entry that inherits both subscriber sets.
Both operations \emph{aim} to preserve recall: each merged subscription represents the union of its inputs' subscribers, so an event matched to a merged subscription notifies all subscribers whose original subscription contributed to the merge. This recall-preservation guarantee is conditional on the matching LLM engaging with the merged subscription set: at high cardinality, the LLM may collapse to a narrow prediction vocabulary and emit empty matches for many events --- a regime we term \emph{empty-prediction collapse} (used hereafter as the canonical name; ``refusal'' appears as descriptive shorthand for an empty structured-output prediction, distinct from the safety-tuning literature's use for explicit task decline) --- even when $|\mathcal{S}|_{\mathrm{eff}}$ fits within the context window (\cref{ssec:crossover}). We therefore refine the discrimination-capacity boundary $|\mathcal{S}|_{\mathrm{cross}}$ to depend on subscription-set \emph{composition} (atomic vs.\ compound merged subscriptions), not cardinality alone; characterising this dependence quantitatively across LLMs and datasets is left to follow-on work.
The subsumption and merge decisions are delegated to the cluster's LLM via the prompt template in Suppl.\ Listing~1: the LLM receives the cluster's subscriptions and identifies which pairs can be covered or merged.
The algorithm applies the LLM's decisions iteratively until a fixed point is reached (no further compression possible).
In principle, \textsc{SubsumptionTest} and \textsc{MergeTest} can also be implemented with deterministic heuristics (e.g., embedding-distance thresholds); the LLM-assisted variant evaluated in this paper delegates both to the same prompt. \emph{Termination.} Every applied cover removes one subscription and every applied merge replaces two subscriptions with one, so each iteration with $\mathit{changed}{=}\textbf{true}$ strictly decreases $|S|$ by at least one while no operation increases it; since $|S|$ is a non-negative integer, \textsc{CoverAndMerge} executes at most $|S|-1$ LLM rounds and is guaranteed to reach a fixed point, independent of the LLM's proposals on any individual round.

\begin{algorithm}
\caption{LLM-assisted subscription compression.}\label{alg:covermerge}
\begin{algorithmic}[1]
\Procedure{CoverAndMerge}{$S,\; \ell$}
    \Repeat
        \State $\mathit{changed} \gets \textbf{false}$
        \State $(\mathit{covers},\, \mathit{merges}) \gets \Call{InvokeLLM}{\ell,\; \text{CoverMergePrompt}(S)}$
        \For{$(s_i, s_j) \in \mathit{covers}$} \Comment{$s_i$ subsumes $s_j$}
            \State $S \gets S \setminus \{s_j\}$;\quad $\mathit{changed} \gets \textbf{true}$
        \EndFor
        \For{$(s_i, s_j) \in \mathit{merges}$}
            \State $S \gets (S \setminus \{s_i, s_j\}) \cup \{\Call{Merge}{s_i, s_j}\}$;\quad $\mathit{changed} \gets \textbf{true}$
        \EndFor
    \Until{$\neg\;\mathit{changed}$}
    \Return $S$ \Comment{Compressed subscription set}
\EndProcedure
\end{algorithmic}
\end{algorithm}

\textbf{Design space.}
The algorithms above expose several configuration knobs that trade off matching accuracy against system cost:
\begin{itemize}
  \item \emph{Number of clusters $k$} controls subscription partitioning granularity. Larger~$k$ yields smaller clusters (less LLM context per invocation, faster prompts) but may split semantically related subscriptions.
  \item \emph{Cover/merge compression} (enabled or disabled) reduces the token footprint of the subscription set at the risk of over-generalisation.
  \item \emph{Reunite flag} ($\mathit{reunite}$, line~8 of \cref{alg:optimize}) merges all compressed clusters back into a single search space after per-cluster compression, maximising LLM context at the cost of larger prompts.
  \item \emph{Cosine threshold $\tau$} controls the precision--recall trade-off of the event-assignment pre-filter. Lower~$\tau$ sends events to more clusters (higher recall, higher cost); higher~$\tau$ restricts assignment.
  \item \emph{Event-side clustering} ($k_e$) optionally pre-groups events before matching, producing fewer but larger batches at the cost of coarser event--subscription granularity.
  \item \emph{Top-$\kappa$ matches} determines how many subscriptions the LLM returns per event.
\end{itemize}
The ablation study in \cref{sec:exp} systematically evaluates the contribution of each component.

\subsection{Prompt Design}\label{ssec:prompts}

The Neural Router uses two prompt templates, reproduced verbatim in Suppl.\ App.~A. The \emph{subscription optimisation prompt} (Listing~1) instructs the LLM to identify which subscriptions within a cluster can be covered or merged; it expects structured JSON output listing the cover/merge decisions consumed by \cref{alg:covermerge}. The \emph{event matching prompt} (Listing~2) presents the compressed subscription set alongside a batch of events and asks the LLM to return the top-$\kappa$ semantically matching subscriptions per event, implementing $\mu$ from \cref{ssec:problem}. Both templates use Python f-string placeholders for runtime injection of cluster subscriptions and event batches.

\subsection{Cost Model}\label{ssec:cost}

Each LLM invocation in the Neural Router consumes a prompt that must fit within the model's context window.
We derive how cover/merge compression affects the dominant cost: the number of LLM invocations during event matching.

\textbf{Context-window constraint.}
Let $W$ denote the context-window size (in tokens), $t_{\mathrm{inst}}$ the fixed token cost of the instruction template, $t_s$ the average token cost of one subscription, $t_e$ the average token cost of one event, and $t_{\mathrm{resp}}$ a token budget reserved for the model's response.
A single matching prompt for cluster~$c$ must satisfy
\begin{equation}\label{eq:window}
t_{\mathrm{inst}} + |c.\mathcal{S}'| \cdot t_s + b \cdot t_e + t_{\mathrm{resp}} \;\leq\; W,
\end{equation}
where $|c.\mathcal{S}'|$ is the number of compressed subscriptions in the cluster and $b$ is the number of events in the batch.
The maximum batch size that fits alongside the subscriptions is therefore
\begin{equation}\label{eq:bmax}
b_{\max}(c) = \left\lfloor \frac{W - t_{\mathrm{inst}} - |c.\mathcal{S}'| \cdot t_s - t_{\mathrm{resp}}}{t_e} \right\rfloor.
\end{equation}

\textbf{Effect of cover/merge compression.}
Let $\rho_c = |c.\mathcal{S}'| / |c.\mathcal{S}| \leq 1$ denote the compression ratio for cluster~$c$ (the fraction of subscriptions retained after cover/merge).
Compression frees $(1 - \rho_c)\,|c.\mathcal{S}|\,t_s$ tokens, increasing the maximum batch size by
\begin{equation}\label{eq:delta}
\Delta b(c) = \left\lfloor \frac{(1 - \rho_c)\,|c.\mathcal{S}|\,t_s}{t_e} \right\rfloor
\end{equation}
additional events per invocation.
This is the primary payoff of compression: it converts subscription token savings into higher event throughput per LLM call.

\textbf{Total matching cost.}
If $m_c$ events are assigned to cluster~$c$ (via the centroid-similarity test in \cref{alg:match}), the number of LLM invocations for that cluster is
\begin{equation}\label{eq:invocations}
I_c = \left\lceil \frac{m_c}{b_{\max}(c)} \right\rceil.
\end{equation}
The total invocations across all active clusters are $I = \sum_{c \in C} I_c$.
With $P$ parallel LLM instances, the number of sequential rounds is $R = \lceil I / P \rceil$.

\textbf{Matching latency.}
Let $\bar{t}_{\mathrm{llm}}$ denote the mean wall-clock time of a single LLM invocation (including prompt encoding, generation, and any network round-trip for cloud-hosted models).
The end-to-end matching latency for a batch of events is
\begin{equation}\label{eq:latency}
L = R \cdot \bar{t}_{\mathrm{llm}} = \left\lceil \frac{I}{P} \right\rceil \cdot \bar{t}_{\mathrm{llm}}.
\end{equation}
When the context window is the binding constraint, compression reduces $I$ (fewer invocations) and batching increases $b_{\max}$ (more events per invocation), both directly reducing $L$.
The per-event amortised latency is $L / m$, which decreases with larger event batches.
For a deployment with $P$ parallel instances and compressed subscriptions, the per-event latency scales as $\mathcal{O}(\bar{t}_{\mathrm{llm}} / (P \cdot b_{\max}))$, making it tuneable through parallelism and compression.

\textbf{Configuration extremes.}
At $|C|{=}1$ (single cluster, $k{=}1$ or $\mathit{reunite}{=}\textbf{true}$), $|c.\mathcal{S}'|$ is large and $b_{\max}$ small: every event matches against all subscriptions, with maximal LLM context. At $|C|{=}k \gg 1$, each cluster carries $|\mathcal{S}'|/k$ subscriptions and $b_{\max}$ rises, but only clusters with queued events incur LLM calls. Event-side clustering ($k_e{>}1$) pre-groups events into larger batches at the cost of coarser event--subscription granularity. The ablation study (\cref{sec:exp}) evaluates these trade-offs.

\textbf{Worked example.}
The benefit of compression depends critically on the context-window size~$W$. Consider a fixed subscription set with $k=10$ clusters, 25 subscriptions each ($|\mathcal{S}|=250$), $m=6{,}000$ events uniform across clusters, and constants $t_{\mathrm{inst}}=200$, $t_s=80$, $t_e=50$, $t_{\mathrm{resp}}=500$. In the \emph{constrained regime} ($W=4$K, typical of edge models), the per-cluster batch size is $b_{\max}=27$ at $\rho=1$, requiring $I=230$ invocations; compression at $\rho=0.6$ frees enough tokens to lift $b_{\max}$ to 43, reducing invocations to $I=140$ (a 39\% reduction). In the \emph{abundant regime} ($W=128$K, typical of cloud APIs), the entire subscription set consumes only a small fraction of the context window: $b_{\max}=2{,}507$ at $\rho=1$ and $2{,}523$ at $\rho=0.6$, both yielding $I=10$ — the 16-event increase from compression is negligible.
Worse, clustering splits subscriptions into $k$ groups, each requiring a separate LLM call: even with only one event per cluster, the minimum invocation count is $k$.
Without clustering (A0), all subscriptions and events fit into 1--2 prompts.
For example, A0 on D1 (19~subscriptions, 100~events) requires only 2 invocations totalling 7{,}514 prompt tokens, whereas A3 (with clustering and compression) requires 24 invocations totalling 11{,}534 prompt tokens.

\textbf{Crossover analysis.}
The cost model predicts a crossover window size $W_{\mathrm{cross}}$ below which compression reduces invocations and above which the clustering overhead dominates. From \cref{eq:bmax}, the per-cluster batch size is $b_{\max}(c) = (W - t_{\mathrm{inst}} - t_{\mathrm{resp}} - t_s |c.\mathcal{S}'|) / t_e$. With $\rho=1$ and $k=1$ (no compression, no clustering) the per-prompt subscription footprint is $t_s|\mathcal{S}|$, while with compression at ratio $\rho$ and $k$ clusters the average per-cluster footprint is $t_s\rho|\mathcal{S}|/k$. Setting $b_{\max}(k=1,\rho=1) \cdot 1 = b_{\max}(k,\rho)\cdot k$ (i.e., the total event-batching capacity matches across configurations) and solving for $W$ yields
\begin{equation}\label{eq:wcross}
W_{\mathrm{cross}} \;=\; t_{\mathrm{inst}} + t_{\mathrm{resp}} + t_s|\mathcal{S}|\cdot\frac{1 - \rho/k}{1 - 1/k}.
\end{equation}
For the per-event token cost, we use the cost-validation production constants $t_s{=}80$, $t_e{=}30$ (\cref{fig:cost-validation}); the worked example above uses an alternative $t_e{=}50$ to illustrate a more conservative event-token estimate (the qualitative picture is unchanged). With $t_{\mathrm{inst}}{=}200$, $t_{\mathrm{resp}}{=}500$, $\rho{=}0.5$, $k{=}9$: for $|\mathcal{S}|{=}19$ (D1), $W_{\mathrm{cross}} \approx 700 + 80\cdot 19 \cdot 0.94 / 0.89 \approx 8\text{K}$ tokens; for $|\mathcal{S}|{=}201$ (D2), $W_{\mathrm{cross}} \approx 700 + 80\cdot 201 \cdot 0.94 / 0.89 \approx 64\text{K}$ tokens.
Modern cloud LLMs operate at $W \geq 128$K (with recent models reaching 200K), well above $W_{\mathrm{cross}}$ for all three evaluation datasets.
This explains the ablation finding that A0 dominates on cost: the context window is no longer the binding constraint, so the overhead of $k$ separate cluster prompts exceeds the savings from compression.
For edge deployments with constrained local models ($W \leq 8$K), the original compression benefit applies in full.

\textbf{Discrimination capacity.}
The crossover analysis above assumes the LLM's matching accuracy is invariant to $|\mathcal{S}|$ as long as all subscriptions fit within the context window. This assumption conflates two resources: \emph{context capacity}~$W$ (how many tokens fit) and \emph{discrimination capacity}~$D$ (how many subscriptions the LLM can evaluate concurrently with acceptable accuracy). The matching prompt requires the LLM to make $|\mathcal{S}|$ simultaneous binary relevance decisions over a shared attention pool, so the per-subscription attention budget is bounded by the model's pool and discrimination degrades as decision load grows. This yields a second crossover point $|\mathcal{S}|_{\mathrm{cross}}$, below which A0 achieves both cost and accuracy optimality, and above which LLM matching accuracy declines --- potentially below independent pairwise baselines (TF-IDF, Sentence-BERT) whose per-pair decisions do not share attention. Crucially, $|\mathcal{S}|_{\mathrm{cross}}$ is model-dependent rather than a global scalar; the empirical evidence and the joint parameter-count $\times$ training-generation factor are reported in \cref{ssec:res-cross,ssec:discrim}.

The two crossover points partition the deployment space into three regimes:
\begin{itemize}
  \item $|\mathcal{S}| < |\mathcal{S}|_{\mathrm{cross}}$ and $W > W_{\mathrm{cross}}$: A0 dominates on both cost and accuracy (D1, D3 across all evaluated models).
  \item $|\mathcal{S}| < |\mathcal{S}|_{\mathrm{cross}}$ and $W < W_{\mathrm{cross}}$: compression pipeline reduces cost while preserving accuracy (edge deployments).
  \item $|\mathcal{S}| > |\mathcal{S}|_{\mathrm{cross}}$: discrimination capacity is the binding constraint regardless of~$W$, but the binding strength is model-dependent. Frontier models may remain viable through partial escape, while resource-constrained models require alternative architectures (e.g., partitioning subscriptions to keep per-prompt $|\mathcal{S}'| \leq D$, or cascade approaches using embedding pre-filters).
\end{itemize}
Full characterisation of the discrimination-capacity scaling law and its dependence on LLM architecture is left to follow-on work; here we identify the boundary qualitatively, validate it empirically at four open-weight data points and one closed-API single-seed reference (\cref{sec:results,sec:discussion}), and observe that the four points are consistent with a joint parameter-count $\times$ training-generation factor rather than with parameter count alone. The single matched-scale cross-family pair (Mistral-7B vs.\ Qwen-2.5-7B) is suggestive but not definitive; broader cross-family validation is required before a closed-form law can be claimed.

\textbf{Token-based cost.}
When backends have different per-token prices, invocation count alone is insufficient. The per-event monetary cost is
\begin{equation}\label{eq:tokencost}
C_{\mathrm{event}} = (T_{\mathrm{prompt}} \cdot p_{\mathrm{in}} + T_{\mathrm{response}} \cdot p_{\mathrm{out}})/m
\end{equation}
where $T_{\mathrm{prompt}}, T_{\mathrm{response}}$ are total prompt/response tokens across all invocations, $p_{\mathrm{in}}, p_{\mathrm{out}}$ are per-token prices, and $m$ is event count. For A0 on D1 ($T_{\mathrm{prompt}}{=}7{,}514$, $T_{\mathrm{response}}{=}2{,}029$, $m{=}100$): Qwen 2.5 (Ollama) is \$0/event (compute amortisation only); Claude 3 Haiku is $\approx$\$0.023/1K; Claude Sonnet is $\approx$\$0.028/1K. The blended-token list-price ratio is approximately $0:1:4$ (Qwen:Haiku:Sonnet), but on this workload Sonnet's prompt and response token counts are not $4{\times}$ Haiku's, so the realised per-event ratio is closer to $0:1:1.2$. Cover/merge offline cost is amortisable: subscription updates occur far less frequently than event arrivals.

\section{Experimental Setup}\label{sec:exp}

We evaluate the Neural Router along eight dimensions:
(1)~a component ablation study isolating the contribution of each design element,
(2)~comparison against embedding-similarity and keyword-based baselines,
(3)~parameter sensitivity analysis,
(4)~validation of the cost model derived in \cref{ssec:cost},
(5)~cross-dataset comparison (varied $|\mathcal{S}|$, modality gap, and capability tier),
(6)~crossover validation under a context-constrained budget (\cref{ssec:crossover-exp}),
(7)~a discrimination-capacity panel sweeping four open-weight backends on D2 (\cref{ssec:discrim-exp}),
and (8)~evaluation of QoE-based heterogeneous backend assignment (\cref{ssec:qoe-exp}).
All configurations are run five times with different random seeds for $k$-means initialisation; we report means and 95\,\% confidence intervals throughout.

\subsection{Datasets}\label{ssec:dataset}

We evaluate on three publicly available datasets (\cref{tab:datasets}), chosen to represent the \emph{content diversity that agents encounter across the computing continuum}: user-generated text at the application layer (D1), governance and regulatory documents (D2), and IoT sensor streams at the device layer (D3). Together, the datasets span a progression of semantic difficulty from keyword-friendly text matching to abductive reasoning over structured sensor data.
In each case, documents or sensor event sequences serve as published \emph{events} and label descriptions serve as natural-language \emph{subscriptions}.
The ground truth defines the matching function: event~$e$ matches subscription~$s$ if and only if $s$ is among $e$'s annotated labels.

\begin{table}[t]
\centering
\caption{Evaluation datasets. $|\mathcal{E}|$\,=\,events, $|\mathcal{S}|$\,=\,subscriptions, $\bar{n}_s$\,=\,mean labels per event, $\bar{w}$\,=\,mean words per event.}
\label{tab:datasets}
\small
\setlength{\tabcolsep}{3.5pt}
\begin{tabular}{@{}lrrrrl@{}}
\toprule
\textbf{Dataset} & $|\mathcal{E}|$ & $|\mathcal{S}|$ & $\bar{n}_s$ & $\bar{w}$ & \textbf{Domain} \\
\midrule
CardiffNLP~\cite{antypas2022twitter} & 6\,000 & 19 & 1.6 & 30 & Social media \\
EUR-Lex~\cite{chalkidis2021multieurlex} & 65\,000 & 201 & 2.2 & 700+ & EU legislation \\
CASAS~\cite{cook2012casas} & 23\,330 & 19 & 1.0 & 45 & Smart home IoT \\
\bottomrule
\end{tabular}
\end{table}

\textbf{D1: CardiffNLP Tweet Topic}~\cite{antypas2022twitter} pairs short English tweets with 19 coarse topics ("sports", "arts \& culture", \ldots).
\textbf{D2: EUR-Lex (MultiEURLEX)}~\cite{chalkidis2021multieurlex} uses the 201 EUROVOC level-2 subject matters as subscriptions; documents are full EU legislative texts averaging 700+ words, stress-testing the context-window constraint (\cref{eq:window}) and the cost model's batch-size predictions.
\textbf{D3: CASAS Smart Home}~\cite{cook2012casas} provides $\approx$\,8 months of free-living sensor traces from home hh113 (PIR motion, magnetic door, temperature), yielding $\approx$\,23\,330 activity segments across 19 activity types (Cook, Eat, Sleep, etc.) after merging fine-grained variants. Each segment is templated into a natural-language event of the form "At \{time\}, sensor in \{location\} reported \{value\}."; subscription descriptions are generic activity descriptions ("Notify when the resident is cooking, involving kitchen activity\ldots") that do not reference sensor identifiers. This stresses the \emph{modality gap}: sensor IDs and binary values (\texttt{M003 ON}, \texttt{D001 OPEN}) share no vocabulary with activity descriptions, and the LLM must perform abductive reasoning to bridge them.

The ablation and baseline comparison (\cref{ssec:ablation,ssec:baselines}) run on all three datasets; the parameter sensitivity sweeps (\cref{ssec:sensitivity}) use A3 on D1--D3 with $k$ matched to the label count where applicable.

\subsection{Ablation Configurations}\label{ssec:ablation}

We define seven configurations that progressively enable each Neural Router component (\cref{tab:ablation-configs}).
Each configuration is evaluated on a primary backend and complemented by a discrimination-capacity sweep on D2 across multiple open-weight scales and one cross-family model:
\begin{itemize}
    \item \textbf{Qwen 2.5 7B} (local, via Ollama): primary open-weight backend, evaluated on D1, D2, D3 for every ablation configuration. Zero API cost; inference latency $\sim$40--50\,s per batch.
    \item \textbf{Qwen 2.5 1.5B, 32B} (local, via Ollama): same-family scaling sweep on D2 to probe how the discrimination-capacity bound depends on parameter count.
    \item \textbf{Mistral 7B} (local, via Ollama): cross-family open-weight model at parameter scale matched to Qwen 2.5 7B, used on D2 to disentangle parameter count from training-generation effects.
    \item \textbf{Claude Sonnet} (Anthropic API): premium commercial backend. Moderate cost; inference latency $\sim$2--5\,s per batch.
    \item \textbf{Claude 3 Haiku} (Anthropic API): lightweight commercial backend. Lowest API cost; inference latency $\sim$1--3\,s per batch.
\end{itemize}
The four open-weight scales (Qwen 1.5B / 7B / 32B and Mistral 7B at matched scale) form the discrimination-capacity panel reported in \cref{ssec:res-cross}; the Qwen 2.5 7B + Anthropic backends carry the main ablation, sensitivity, and scaling tables.
For Sonnet, we use a single seed to control API cost. For D2 with Sonnet, we additionally cap at 5\,000 stratified events (7.7\,\% of the full corpus), since the combination of long legal documents and a commercial API model at scale would be prohibitively expensive. For the Qwen-2.5-7B runs executed on a shared HPC partition with a 3-hour per-task walltime budget, events are stratified-subsampled to 1\,000 (D1, D3) and 300 (D2). This is a compute-budget constraint, not an architectural one, and is uniform across configurations so within-dataset comparisons remain apples-to-apples. The Haiku full-corpus evaluation establishes that the qualitative findings hold at full scale on the same dataset.
Configurations A5 (event clustering) and A6 (no cosine filter) are evaluated on D1 and D3 only; on D2 the discrimination-capacity finding (\cref{ssec:res-cross}) implies all configurations struggle equally because the binding constraint is the LLM's per-subscription attention budget, so adding A5 and A6 measurements on D2 contributes negligible insight beyond what A0--A4 already establish.
All configurations use Sentence-BERT (all-MiniLM-L6-v2) as the default embedding model unless otherwise noted in the sensitivity analysis (\cref{ssec:sensitivity}). The experiment pipeline is managed by DVC~\cite{dvc2020} for full reproducibility; all stages, parameters, and intermediate results are version-controlled.

\textbf{LLM decoding and reproducibility.}
All LLM calls use temperature\,=\,0 (greedy decoding) to minimise non-determinism. The matching and cover/merge prompts request structured JSON output, which is parsed deterministically. Despite greedy decoding, exact reproducibility is not guaranteed across LLM invocations: cloud API models may be updated by the provider, and local GPU inference can exhibit floating-point non-determinism across hardware or driver versions. We mitigate this by recording the API model snapshot identifiers (see \cref{ssec:environment}) and by reporting results over multiple seeds, so that the variance attributable to $k$-means initialisation is captured explicitly. The five-seed protocol primarily controls for clustering randomness; LLM output variation at temperature\,=\,0 is empirically negligible within a single model snapshot.

\textbf{Seed and subsample protocol.} The ablation, cross-dataset, and discrimination-capacity tables use five $k$-means seeds $\{42, 123, 456, 789, 1024\}$ for every (dataset, configuration) cell on the open-weight Qwen and Mistral backends; Sonnet rows are single-seed by API-cost necessity and are marked indicative. The QoE perturbation (\cref{ssec:res-qoe-pert}) and calibration-fraction (\cref{ssec:res-qoe-calfrac}) experiments use $n{=}15$ seeds under a matched-pair regime with an on-disk LLM-call cache, so GPU and decoding nondeterminism cancel in perturbed-vs-baseline deltas; the wider-tier 7B/72B companion (Suppl.\ App.~F) likewise uses $n{=}15$. Subsample caps ($1{,}000$ events for D1/D3 and $300$ for D2 on Qwen-2.5-7B; $5{,}000$ for Sonnet on D2) are compute-budget constraints applied uniformly within a dataset, so within-dataset comparisons remain matched-pair. Per-cell run provenance --- SLURM job identifiers, completion timestamps, and regenerated-CSV paths --- is recorded in the artefact's operations log; the one dropped baseline-vs-injection cell in \cref{ssec:res-qoe-pert} ($n{=}14$ for that comparison) is noted in place.

\begin{table}[t]
\centering
\footnotesize
\caption{Ablation configurations. Default parameters: $k{=}19$, $\tau{=}0.3$, $\kappa{=}3$.}
\label{tab:ablation-configs}
\setlength{\tabcolsep}{4pt}
\begin{tabular}{@{}lccccc@{}}
\toprule
\textbf{Config} & \rotatebox{70}{\textbf{Cluster}} & \rotatebox{70}{\textbf{C\&M}} & \rotatebox{70}{\textbf{Reunite}} & \rotatebox{70}{\textbf{Evt.\ clust.}} & \rotatebox{70}{\textbf{Cos.\ filter}} \\
\midrule
A0: Raw LLM        & --  & --  & --  & --  & --  \\
A1: Cluster only    & \checkmark & --  & --  & --  & \checkmark \\
A2: C\&M only       & --  & \checkmark & --  & --  & --  \\
A3: Clust.\ + C\&M  & \checkmark & \checkmark & --  & --  & \checkmark \\
A4: + Reunite       & \checkmark & \checkmark & \checkmark & --  & --  \\
A5: + Evt.\ clust.  & \checkmark & \checkmark & --  & \checkmark & \checkmark \\
A6: No cos.\ filter & \checkmark & \checkmark & --  & --  & --  \\
\bottomrule
\end{tabular}
\end{table}

\Cref{tab:ablation-configs} shows the seven configurations. \textbf{A0} batches all subscriptions and events into a single LLM prompt with no clustering or compression --- the accuracy ceiling and cost floor below the discrimination-capacity crossover. \textbf{A1} clusters subscriptions into $k$ groups and routes events by centroid cosine similarity ($\tau$) without cover/merge. \textbf{A2} applies cover/merge to the full unpartitioned subscription set. \textbf{A3} enables both clustering and compression. \textbf{A4} adds the $\mathit{reunite}$ flag, merging compressed clusters back into a single search space. \textbf{A5} adds event-side clustering ($k_e{=}5$) on top of A3. \textbf{A6} sets $\tau{=}0$ (all events to all clusters), isolating the cosine pre-filter's contribution.

\subsection{Baselines}\label{ssec:baselines}

We compare against seven baselines spanning keyword matching, embedding similarity, pairwise neural scoring, and zero-shot classification: BM25~\cite{robertson2009bm25} as the canonical keyword baseline; Sentence-BERT cosine using the same encoder as the Neural Router (\texttt{all-MiniLM-L6-v2}), isolating the LLM's contribution beyond embedding geometry; a cross-encoder reranker (\texttt{cross-encoder/ms-marco-MiniLM-L-6-v2}) as a pairwise neural scoring model; zero-shot classification with DistilBART-MNLI-12-1 \cite{lewis2020bart,yin2019benchmarking} \\(\texttt{valhalla/distilbart-mnli-12-1}, distilled from \texttt{facebook/bart-large-mnli}; events classified against subscription descriptions as candidate labels via natural language inference~\cite{williams2018broad,wang2018glue,laurer2024less}); and GloVe, TF-IDF, Word2Vec as static embedding baselines retained for continuity. All baselines return a ranked subscription list per event and are evaluated at the same top-$\kappa$ cut-off as the Neural Router.

\subsection{Parameter Sensitivity}\label{ssec:sensitivity}

We sweep four parameters, varying one at a time with the others fixed at the A3 defaults: cluster count $k \in \{1, 5, 10, 15, 19, 25, 30\}$ ($k{=}19$ matches the ground-truth topic count); cosine threshold $\tau \in \{0.0, 0.1, \ldots, 0.9\}$; top-$\kappa$ matches $\kappa \in \{1, 2, 3, 5, 7, 10\}$; and embedding model (all-MiniLM-L6-v2, all-mpnet-base-v2, e5-large-v2, bge-base-en-v1.5).

\subsection{Cost Model Validation}\label{ssec:cost-validation}

For every configuration in \cref{ssec:ablation,ssec:sensitivity}, we record per-cluster compression ratio $\rho_c$, measured invocations $I_{\mathrm{meas}}$, prompt + response tokens per invocation, and per-step wall-clock times (embedding, clustering, cover/merge, cosine filtering, prompt packing, LLM inference, post-processing); we compare $I_{\mathrm{meas}}$ against the predicted $I_{\mathrm{pred}}$ from \cref{eq:invocations} (\cref{fig:cost-validation}). Cost is reported as tokens (prompt + response) per event; per-event monetary cost follows \cref{eq:tokencost} with Qwen 2.5 7B at zero API cost, Claude 3 Haiku at \$0.80/M input + \$4.00/M output, and Claude Sonnet at \$3.00/M input + \$15.00/M output. At blended-token list price the per-token cost ratio is approximately $0:1:4$ (Qwen:Haiku:Sonnet); the realised per-event cost ratio in \cref{tab:ablation} is closer to $0:1:1.2$ on D1 because Sonnet's prompt and response token counts are not $4\times$ Haiku's on this workload.

\subsection{Scaling Analysis}\label{ssec:scaling}

We evaluate how the system behaves as the subscription set grows: empirically by subsampling and duplicating subscriptions to size $|\mathcal{S}| \in \{50, 100, 250, 500, 1000, 2000\}$ and measuring $I$, $L$, $\rho$, F1 for A3; analytically by projecting $I$ and $L$ for $|\mathcal{S}| \in \{5{,}000, 10{,}000, 50{,}000, 100{,}000\}$ from the validated cost model with measured $\rho$ and $\bar{t}_{\mathrm{llm}}$ (\cref{fig:scaling}).

\subsection{Crossover Validation}\label{ssec:crossover-exp}

To validate the context-window crossover predicted in \cref{ssec:cost}, we enforce $W=4{,}096$ tokens on the matching prompt and sweep $|\mathcal{S}| \in \{50, 200, 2{,}000\}$ on D1 via subsampling and duplication-with-rename, comparing A0 (raw LLM, list truncated to fit) against A4 (full pipeline with \textsc{CoverAndMerge}). The budget is enforced by truncating the in-prompt subscription list before invoking the LLM, isolating the prediction from model-capacity confounds. We report F1, end-to-end latency, and per-event token cost; the crossover point is where A4 first matches or exceeds A0 in F1. This is distinct from the discrimination-capacity crossover $|\mathcal{S}|_{\mathrm{cross}}$, which concerns accuracy at large $|\mathcal{S}|$ regardless of $W$. Single seed on Qwen-2.5-7B at temperature 0; backends with native $W$ ($\geq 32$\,K for local Qwen, $\geq 128$\,K for cloud APIs) are not context-constrained at $|\mathcal{S}| \leq 2{,}000$ and serve as upper-bound references.

\subsection{Discrimination-Capacity Panel}\label{ssec:discrim-exp}

The cost model's discrimination-capacity crossover (\cref{ssec:cost}) predicts that LLM matching accuracy declines when the subscription cardinality $|\mathcal{S}|$ approaches the model's effective discrimination capacity. To probe how the boundary depends on the LLM, we evaluate four open-weight models on D2 (the dataset with $|\mathcal{S}|=201$, where the boundary is most consequential) under matched experimental conditions:

\begin{itemize}
  \item \textbf{Same-family parameter sweep.} Qwen 2.5 at three scales --- 1.5B, 7B, 32B --- with all other factors held constant. This isolates parameter count as the independent variable within a single training generation and tokenizer.
  \item \textbf{Cross-family control at matched scale.} Mistral 7B Instruct at the same parameter count as the middle Qwen point, with the same prompt and decoding settings. Comparing Qwen 7B and Mistral 7B isolates training generation / instruction-tuning sophistication at fixed parameter count.
\end{itemize}

All four models are evaluated on configurations A0 (raw LLM) and A1 (clustering only), which together span the relevant points on the cost-accuracy frontier on D2 (per the \cref{ssec:res-ablation} ablation, A0 and A1 are the strongest configurations on D2 under the existing Qwen 7B baseline). Five seeds each, MAX\_EVENTS$=$300 stratified subsample (matching the Qwen 7B D2 cap); 40 runs total. Where useful, the existing Haiku full-corpus and Sonnet 5{,}000-event subsample numbers are reported as upper-bound references against the open-weight panel.

\subsection{QoE Heterogeneous Backend Assignment}\label{ssec:qoe-exp}

To explore the QoE-based backend assignment (\cref{eq:qoe}), we evaluate three assignment strategies on D1 with a Qwen-2.5 tier gradient (7B, 32B): homogeneous (all clusters to a single backend, the ablation default), round-robin (clusters cyclically assigned), and QoE-optimised (each cluster to the backend maximising $\mathit{QoE}(c, \ell)$ after the calibration phase of \cref{sec:design}). Calibration uses 10\% of events per cluster per backend; the remaining 90\% are used for evaluation. Three normalised weight configurations (sum to one) explore the accuracy-cost Pareto: accuracy-first $(0.70, 0.15, 0.15)$, balanced $(0.34, 0.33, 0.33)$, and cost-first $(0.15, 0.70, 0.15)$. We additionally exercise the loop with two perturbations (topic-restricted calibration; latency injection) and a calibration-fraction sweep $\in \{0.05, 0.10, 0.20, 0.50, 0.80, 1.00\}$ under matched-pair semantics, with all cells of a sweep sharing one SLURM job and an on-disk LLM-call cache keyed on $(\mathrm{model}, \mathrm{prompt})$ so GPU and decoding nondeterminism cancel in the perturbed-vs-baseline delta. Results in \cref{ssec:res-qoe,ssec:res-qoe-pert,ssec:res-qoe-calfrac}.

\subsection{Metrics}\label{ssec:metrics}

\textbf{Matching accuracy:} precision, recall, F1 score, and false positive rate (FPR), computed per-event and macro-averaged. Let $E$ denote the event corpus, $\mathcal{S}$ the subscription set, $G(e)\subseteq\mathcal{S}$ the ground-truth subscription IDs for event $e$, and $P(e)\subseteq\mathcal{S}$ the predicted IDs returned by the matcher. ID-based per-event precision and recall are $\mathrm{Prec}_{\mathrm{id}}(e)=|P(e)\cap G(e)|/|P(e)|$ and $\mathrm{Rec}_{\mathrm{id}}(e)=|P(e)\cap G(e)|/|G(e)|$; macro-averaged F1 is the harmonic mean per event averaged over $e\in E$. A match is correct iff the returned subscription ID belongs to the event's ground-truth set.

\textbf{Description-aware F1.}\label{def:desc-f1} When the subscription set contains multiple distinct IDs sharing one description (the duplication-with-rename construction used in \cref{ssec:crossover} to test $|\mathcal{S}|>|\mathcal{S}|_\mathrm{native}$), ID-based F1 mis-counts a CoverAndMerge correct merging of near-duplicates as missed matches. Let $d:\mathcal{S}\to\mathcal{D}$ map each ID to its free-text description, and let $\bar{P}(e)=\{d(s):s\in P(e)\}$ and $\bar{G}(e)=\{d(s):s\in G(e)\}$ be the description-sets of prediction and ground truth. \emph{Description-aware} per-event precision and recall are
\[
\mathrm{Prec}_{\mathrm{desc}}(e)=\frac{|\bar{P}(e)\cap \bar{G}(e)|}{|\bar{P}(e)|},\qquad \mathrm{Rec}_{\mathrm{desc}}(e)=\frac{|\bar{P}(e)\cap \bar{G}(e)|}{|\bar{G}(e)|},
\]
with the macro-averaged F1 defined identically to the ID-based variant. Compound merged IDs of the form $\texttt{id}_i\texttt{+}\texttt{id}_j$ produced by CoverAndMerge are split on \texttt{+} before applying $d$, so a merged prediction matches against any of the original descriptions it represents. \emph{Reduction-invariant statement:} when every ID has a distinct description (i.e., $d$ is injective), description-aware F1 reduces to ID-based F1; the two metrics differ only on duplication-with-rename subscription sets. The synthetic-data invariant test \texttt{tests/test\_synthetic\_data\_metric\_invariant.py} in the artefact repository asserts this reduction together with the round-trip property "ID-based F1 underestimates description-aware F1 on duplicated sets".

\textbf{System performance:} end-to-end matching latency $L$ (\cref{eq:latency}), per-event amortised latency $L/m$, throughput (events/second), and LLM invocation count $I$.

\textbf{Cost:} compression ratio $\rho$, tokens consumed, and monetary cost per 1\,000 events.

\textbf{Statistical reporting:} all metrics are reported as mean $\pm$ 95\,\% CI over 5 seeds. Pairwise significance is assessed via the Wilcoxon signed-rank test; with $n{=}5$ paired observations the minimum attainable two-sided $p$ is $0.0625$, so directional consistency across seeds is the primary evidence and the $\alpha{=}0.05$ rejection bar is reported for completeness only.

\textbf{Reproducibility apparatus.} Five regression invariants in the artefact repository gate figure rendering against the headline claims: I1--I4 cover the F1, empty-prediction-rate, and cost-model claims; \textbf{I5} asserts calibration--evaluation event-set disjointness, $E_{\mathrm{cal}}\cap E_{\mathrm{eval}}=\emptyset$ per (cluster, backend), which is load-bearing for the perturbation and calibration-fraction experiments of \cref{ssec:res-qoe-pert,ssec:res-qoe-calfrac} (the sole exception is the $\mathrm{frac}{=}1.00$ asymptotic-ceiling point, where disjointness is intentionally not imposed and the result is read as an upper bound). The invariants run as a blocking \texttt{pytest} stage in the DVC pipeline (\texttt{dvc.yaml} stage \texttt{regression-invariants}): a failing invariant fails the pipeline and blocks figure regeneration, so figures cannot drift from claims (Appendix~G in the supplementary material).

\subsection{Implementation and Environment}\label{ssec:environment}

The Neural Router is implemented in Python~3.10 with embeddings via sentence-transformers (all-MiniLM-L6-v2) and clustering via scikit-learn. LLM inference uses six backends via LiteLLM: \emph{Qwen 2.5 1.5B / 7B / 32B} and \emph{Mistral 7B Instruct} run locally via Ollama at the default \texttt{Q4\_K\_M} 4-bit GGUF quantisation (digests in \texttt{ollama-digests.txt}); \emph{Claude 3 Haiku} (\texttt{claude-3-haiku-20240307}) and \emph{Claude Sonnet} (\texttt{claude-sonnet-4-20250514}) run via the Anthropic API EU endpoint. The 7B Qwen variant is the primary open-weight backend across all ablations. Open-weight inference runs on CSC Puhti (NVIDIA V100 16\,GB) for the main ablation table and CSC Mahti (NVIDIA A100 40\,GB) for the cost-model validation re-run with per-cluster instrumentation (the additional logging needs the larger card's memory) and for the 1.5B/32B Qwen and Mistral runs. Cross-hardware F1 was not separately validated; temperature-0 accuracy is FP-deterministic up to rounding noise. Reported $\bar{t}_{\mathrm{llm}}$ includes cloud API network round-trip for Haiku and Sonnet; local backends do not. Cloud model outputs may vary across provider updates, so results are indicative for the current model generation rather than exact reproducibility targets. Source code, experiment scripts, DVC pipeline definitions, regression tests (including the metric-invariant tests cited in \cref{ssec:metrics}), and the per-event prediction parquet files for the \cref{ssec:crossover} sweep are at \selfurl{}.

\section{Results}\label{sec:results}

\subsection{Ablation Study and Baseline Comparison}\label{ssec:res-ablation}

\Cref{tab:ablation} reports matching accuracy and system cost on D1 for the seven ablation configurations defined in \cref{ssec:ablation} alongside the seven model-independent baselines (\cref{ssec:baselines}); D2 and D3 ablation results appear in Suppl.\ Tab.~1, and the D2 open-weight discrimination-capacity panel in \cref{tab:d2-discrim}.

Three patterns are visible. \emph{(i)~A0 and A4 dominate the ablation frontier:} A0 (raw LLM) attains $\mathrm{F1}{=}0.656_{\pm 0.001}$ at $I{=}450$ invocations, and A4 (full pipeline with \textsc{CoverAndMerge}) reaches $0.638_{\pm 0.019}$ at $I{=}445$, recovering nearly all of A0's accuracy at the same invocation budget. \emph{(ii)~Cluster-only variants (A1, A3, A5) trade an order of magnitude more invocations ($\sim 4{,}600{-}4{,}700$) for $\mathrm{F1}\approx 0.44$,} a strictly worse cost--accuracy point; A6 (no cosine filter) is dominated on every column. \emph{(iii)~Every Neural Router configuration except A2 and A6 outperforms the strongest non-LLM baseline} (DistilBART-MNLI zero-shot, $\mathrm{F1}{=}0.434$) by $\geq 0.01$ absolute F1, and A0/A4 do so by $\geq 0.20$ absolute F1; the four cosine-baseline rows (Sentence-BERT, GloVe, TF-IDF, Word2Vec) span $\mathrm{F1}\in[0.10,0.42]$, confirming the semantic-matching gap that motivates the LLM-driven design.

\begin{table*}[t]
\centering
\footnotesize
\caption{Ablation and baseline results on D1 (CardiffNLP, full 6{,}000-event corpus). Backend: Claude Haiku; F1 reported as $\mathrm{mean}_{\pm \mathrm{half\text{-}CI95}}$ over 5 seeds, $k{=}19$, $\kappa{=}3$. Best Neural Router F1 in \textbf{bold}. $I$\,=\,LLM invocations, $\rho$\,=\,compression ratio (n/a for baselines and uncompressed configs reported as 1.00), $L$\,=\,end-to-end latency. Cost column from \cref{ssec:cost} worked example at Haiku rates. Results for D2 and D3 in Suppl.\ Tab.~1; the D2 \emph{open-weight discrimination-capacity panel} appears in \cref{tab:d2-discrim}.}
\label{tab:ablation}
\setlength{\tabcolsep}{4.5pt}
\begin{tabular}{@{}lcccccccc@{}}
\toprule
\textbf{Configuration} & \textbf{Precision} & \textbf{Recall} & \textbf{F1} & \textbf{FPR} & $\boldsymbol{I}$ & $\boldsymbol{\rho}$ & $\boldsymbol{L}$\textbf{\,(s)} & \textbf{\$/1k evt} \\
\midrule
\multicolumn{9}{@{}l}{\textit{Neural Router ablation (Claude Haiku, D1)}} \\
\addlinespace[2pt]
A0: Raw LLM           & 0.712 & 0.674 & \textbf{$0.656_{\pm 0.001}$}      & 0.030 &   450 & 1.00 &  3.1 & 0.023 \\
A1: Cluster only      & 0.552 & 0.398 & $0.443_{\pm 0.000}$               & 0.018 & 4{,}679 & 1.00 &  5.4 & 0.18 \\
A2: C\&M only         & 0.308 & 0.523 & $0.361_{\pm 0.080}$               & 0.083 &   382 & 1.24 &  3.5 & 0.020 \\
A3: Clust.\ + C\&M    & 0.551 & 0.397 & $0.442_{\pm 0.002}$               & 0.018 & 4{,}672 & 1.00 &  4.7 & 0.18 \\
A4: + Reunite         & 0.683 & 0.667 & $0.638_{\pm 0.019}$               & 0.033 &   445 & 1.00 &  3.2 & 0.023 \\
A5: + Evt.\ clust.    & 0.547 & 0.395 & $0.439_{\pm 0.004}$               & 0.018 & 4{,}637 & 1.00 &  5.2 & 0.18 \\
A6: No cosine filter  & 0.179 & 0.312 & $0.216_{\pm 0.038}$               & 0.131 & 7{,}751 & 1.00 & 16.2 & 0.31 \\
\midrule
\multicolumn{9}{@{}l}{\textit{Baselines (model-independent; D1)}} \\
\addlinespace[2pt]
BM25                       & 0.070 & 0.114 & 0.082 & 0.161 & 0  & n/a &      1.8 & 0 \\
Sentence-BERT cosine       & 0.334 & 0.656 & 0.423 & 0.115 & 0  & n/a &     21.5 & 0 \\
Cross-encoder              & 0.258 & 0.491 & 0.323 & 0.128 & 0  & n/a &    433.1 & 0 \\
DistilBART-MNLI zero-shot  & 0.340 & 0.684 & 0.434 & 0.114 & 0  & n/a & 10{,}323 & 0 \\
GloVe cosine               & 0.084 & 0.153 & 0.104 & 0.158 & 0  & n/a &      0.1 & 0 \\
TF-IDF cosine              & 0.084 & 0.147 & 0.102 & 0.158 & 0  & n/a &      0.2 & 0 \\
Word2Vec cosine            & 0.090 & 0.161 & 0.110 & 0.157 & 0  & n/a &      0.1 & 0 \\
\bottomrule
\end{tabular}
\end{table*}

\subsection{Parameter Sensitivity}\label{ssec:res-sensitivity}

We sweep four pipeline hyperparameters one at a time on D1 with the Qwen-2.5-7B backend in the A3 (full pipeline) configuration; Suppl.\ Fig.~1 reports the four sweeps as a single multi-panel figure with panels (a)--(d). Panel~(a), the cluster count $k$, gives a non-monotone F1 response: $k{=}1$ (degenerate single-cluster) and $k{=}19$ (one cluster per native subscription) both attain higher F1 than the intermediate values $k \in \{5, 10, 15\}$ at near-constant invocation cost. We retain $k{=}19$ because it cleanly separates clusters per topic, yielding the lowest false-positive rate. Panel~(b), the cosine threshold $\tau$, shows F1 plateauing above $\tau \approx 0.3$ as the centroid-fallback path dominates; the effective operating range is $\tau \in [0.05, 0.2]$, where lower $\tau$ buys recall at the cost of additional invocations (at $\tau{=}0$, every event goes to every cluster, $I{=}301$ vs.\ $I{=}31$ at $\tau{=}0.3$). Panel~(c) shows that $\kappa{=}3$ is at the F1 plateau (matching $\kappa{=}5$ within rounding); we adopt $\kappa{=}3$ as the most parsimonious choice. Panel~(d), the embedding-model response, spans F1 $\in [0.263, 0.300]$ across four mainstream sentence-transformer encoders; the larger e5-large-v2 and BGE-base attain the highest F1 but at $\sim 10\times$ the invocation count (their cosine-similarity distributions concentrate the assignment fallback through more clusters); we retain MiniLM-L6 as the lightweight default because it sits within $0.04$ F1 of the best at $\sim 1/12$ the cost.

\subsection{Cost-Model and Scaling Validation}\label{ssec:res-cost}

\begin{figure}[t]
  \centering
  \begin{subfigure}[t]{0.32\linewidth}
    \centering
    \includegraphics[width=\linewidth]{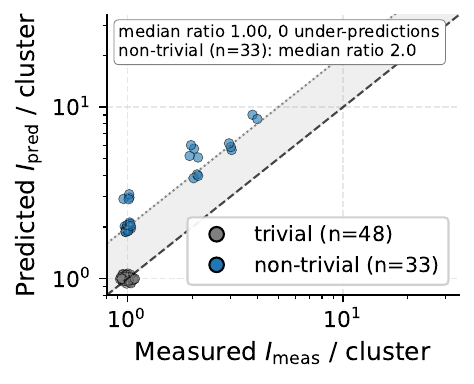}
    \caption{Cost-model validation: predicted vs.\ measured per-cluster invocations on D1.}
    \label{fig:cost-validation}
  \end{subfigure}\hfill
  \begin{subfigure}[t]{0.32\linewidth}
    \centering
    \includegraphics[width=\linewidth]{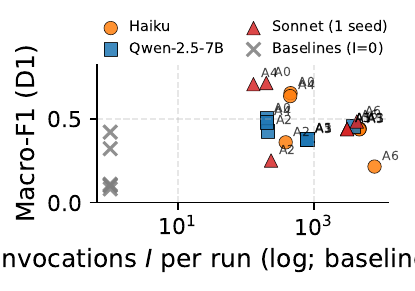}
    \caption{Accuracy--cost trade-off on D1 across three backends and seven configurations.}
    \label{fig:pareto}
  \end{subfigure}\hfill
  \begin{subfigure}[t]{0.32\linewidth}
    \centering
    \includegraphics[width=\linewidth]{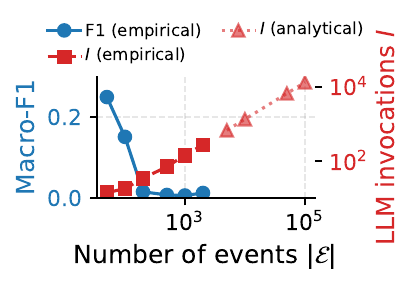}
    \caption{Event-count scaling on D3 with Qwen-2.5-7B in configuration A3.}
    \label{fig:scaling}
  \end{subfigure}
  \caption{Cost-model and scaling validation. \textbf{(\subref{fig:cost-validation})} Predicted vs.\ measured per-cluster LLM invocations across configs $\{$A0, A1, A3$\}$ and $k \in \{1, 2, 5, 10, 19\}$, $n{=}81$ markers (Qwen-2.5-7B on Mahti GPU and dry-run client). Predicted $I_{\mathrm{pred}}$ from \cref{eq:invocations} with production constants. The model is a \emph{conservative ceiling} (median ratio $1.00$, \emph{zero} under-predictions): the grey wedge marks the one-sided within-$2\times$ region, trivial cells ($m_c{\le}b_{\max}$, $n{=}48$) lie exactly on the line, and non-trivial cells ($n{=}33$, median ratio $2.00$) over-predict, $58\%$ within $2\times$; $83\%$ in-band overall. Stratified detail in body text. Both axes log. \textbf{(\subref{fig:pareto})} F1 vs.\ invocation count $I$ (log) on D1; embedding baselines plotted at $I{=}1$. Within each backend, A0 and A4 attain the best F1--$I$ point and A6 the highest $I$. \textbf{(\subref{fig:scaling})} F1 (left) and $I$ (right, log) for $|\mathcal{E}| \in \{50, ..., 2{,}000\}$ at fixed $\sim$5--7 events/call; F1 declines as the per-event budget shrinks (\cref{ssec:cost}). Dotted points project $I$ to $|\mathcal{E}| \in \{5{,}000, ..., 100{,}000\}$ from the linear fit.}
  \label{fig:cost-panel}
\end{figure}

\Cref{fig:cost-panel} validates the cost model along three views. Panel \subref{fig:cost-validation} compares predicted to measured per-cluster invocations on D1 (one marker per $(config, k)$ cell, with $I_{\mathrm{pred}}{=}\lceil m_c / b_{\max}(c) \rceil$): the median predicted/measured ratio is $1.00$ and $83\%$ of the $n{=}81$ cells fall within the factor-of-two band. \emph{Stratified miss analysis} (per \texttt{analysis/cost\_validation\_stratified.py}): all band-violations fall in the $n{=}33$ non-trivial cells ($m_c{>}b_{\max}$, where the ceiling is genuinely under test: $14$ misses, $42\%$, median ratio $2.00$), while the $n{=}48$ trivial cells ($m_c{\le}b_{\max}$) are exact and the model never under-predicts; the $83\%$-in-band headline thus aggregates an exact trivial stratum with a weaker $58\%$-in-band non-trivial fit (\cref{ssec:limitations}). Panel \subref{fig:pareto} plots accuracy versus cost across all $(config, backend)$ cells on D1: each backend's A0 and A4 attain the best F1 at the lowest invocation count; A6 (no cosine filter) is the most expensive configuration; A1/A3/A5 (multi-cluster routing) sit at moderate-to-high cost; embedding baselines occupy the no-LLM band at the left edge. Panel \subref{fig:scaling} sweeps event count $|\mathcal{E}| \in \{50, ..., 2{,}000\}$ on D3 in configuration A3: F1 declines from $0.25$ to below $0.02$ as the per-event share of the 4{,}096-token budget shrinks (consistent with the discrimination-capacity argument of \cref{ssec:cost}), while $I$ grows linearly ($R^2 \geq 0.99$); the dotted right-axis points project $I$ to $|\mathcal{E}| \in \{5{,}000, ..., 100{,}000\}$ from the linear fit.

\textbf{Latency breakdown.}
The latency decomposition reveals that LLM inference dominates end-to-end matching time, accounting for $>$95\,\% of total latency across all configurations. Embedding ($<$0.5\,s for 11K events), clustering ($<$0.1\,s), and cosine pre-filtering ($<$0.01\,s) are negligible. The practical implication is that latency optimisation should focus on LLM selection and batching (via the QoE-based cost model of \cref{ssec:cost}), not on the embedding or clustering pipeline.

\subsection{Crossover Validation}\label{ssec:crossover}

\begin{figure}[t]
  \centering
  \begin{subfigure}[t]{0.58\linewidth}
    \centering
    \includegraphics[width=\linewidth]{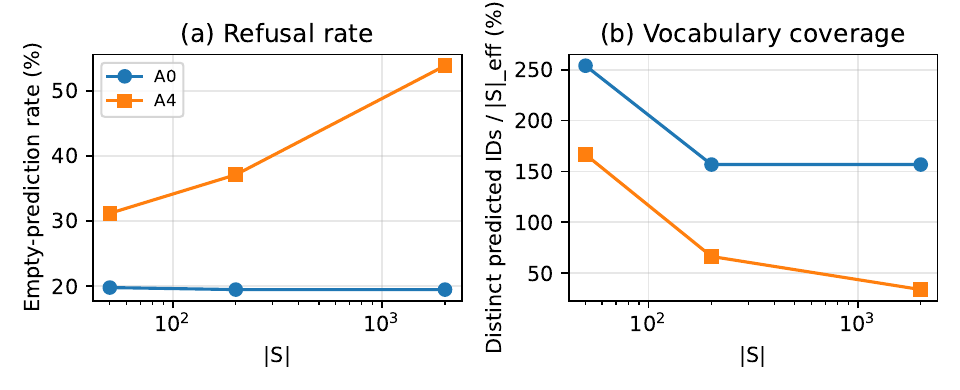}
    \caption{Per-event mechanism: empty-prediction rate (left) and distinct-prediction vocabulary as a fraction of $|\mathcal{S}|_{\mathrm{eff}}$ (right) for A0 and A4 across $|\mathcal{S}|$.}
    \label{fig:crossover-mechanism}
  \end{subfigure}\hfill
  \begin{subfigure}[t]{0.38\linewidth}
    \centering
    \includegraphics[width=\linewidth]{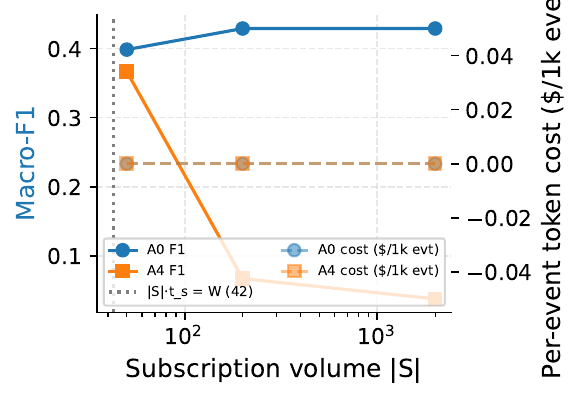}
    \caption{Aggregate F1 (left) and per-event token cost (right) across the same $|\mathcal{S}|$ sweep.}
    \label{fig:crossover}
  \end{subfigure}
  \caption{Empirical crossover validation on D1 with Qwen-2.5-7B at $W{=}4{,}096$, description-aware F1, single seed (subsampling and duplication-with-rename, evaluated under description-aware F1 to neutralise the duplication-with-rename metric-invariant). Panel (\subref{fig:crossover-mechanism}) shows the per-event signature explaining panel (\subref{fig:crossover})'s aggregate gap as empty-prediction collapse and vocabulary narrowing rather than mid-prediction compound-description disambiguation; the vertical dotted line in (\subref{fig:crossover}) marks the analytical threshold $|\mathcal{S}|\!\cdot\!t_s = W$. Numerical detail in the body text.}
  \label{fig:crossover-panel}
\end{figure}

The cost model (\cref{ssec:cost}) predicts that under a constrained context window, the compression pipeline (A4) preserves accuracy where raw prompting (A0) must drop subscriptions. We test this prediction on D1 with Qwen-2.5-7B at $W{=}4{,}096$ tokens, sweeping $|\mathcal{S}| \in \{50, 200, 2{,}000\}$ via subsampling and duplication-with-rename, single seed (\cref{fig:crossover}). Because duplication-with-rename produces multiple subscription IDs sharing one description, an ID-based F1 mis-counts CoverAndMerge's correct merging of near-duplicates as missed matches. We therefore evaluate this experiment under the \emph{description-aware} F1 defined formally in \cref{ssec:metrics}, which collapses both predictions and ground truth to description sets before computing precision and recall, so semantically equivalent merges count as hits.\footnote{An earlier version of this experiment used ID-based F1 and produced an apparent collapse that was traced to this metric-invariant violation; the description-aware run reported here is the corrected measurement.}

\emph{For A0}, F1 plateaus at $0.43$ once $|\mathcal{S}| \geq 200$: the $W{=}4{,}096$ truncation caps the in-prompt subscription count at $\approx 162$, and matching against any $162$-subscription subset of D1 yields the same accuracy. \emph{For A4}, F1 \emph{decreases} with $|\mathcal{S}|$ ($0.367 \to 0.067 \to 0.038$) where the cost model predicts increase. The decline is not a metric artifact: A4's CoverAndMerge correctly compresses the duplicated set ($\rho = 0.6$ at $|\mathcal{S}|=50$, $\rho = 0.42$ at $|\mathcal{S}|=2{,}000$), and the merged set's effective size $|\mathcal{S}|_{\mathrm{eff}}$ stays in the $125$--$133$ range across the high-$|\mathcal{S}|$ cells, comparable to A0's $|\mathcal{S}|_{\mathrm{eff}}{=}162$. Per-event analysis (\cref{fig:crossover-mechanism}) reveals the failure mode: as $|\mathcal{S}|$ grows under A4, the LLM (i) emits an empty match for a growing fraction of events (refusal rate $19\% \to 31\% \to 54\%$) and (ii) draws its non-empty predictions from a shrinking subset of the active subscription set ($167\% \to 66\% \to 34\%$ of $|\mathcal{S}|_{\mathrm{eff}}$, where values above $100\%$ indicate the LLM also hallucinates IDs not in the set). Merged subscriptions with compound IDs are picked $38$ times at $|\mathcal{S}|{=}50$ and \emph{zero} times at $|\mathcal{S}|{=}2{,}000$. A0's per-event statistics stay flat across $|\mathcal{S}|$ ($\approx 20\%$ refusal, $\approx 160\%$ vocabulary used). The mechanism is therefore consistent with \emph{empty-prediction collapse and vocabulary narrowing} (we use ``refusal'' descriptively for the empty structured-output predictions, distinct from the safety-tuning literature's use for explicit task decline): faced with a subscription set that mixes atomic and compound merged subs at high cardinality, Qwen-2.5-7B narrows its prediction vocabulary and increasingly returns empty matches. Cross-seed reproducibility is not formally tested at this single seed; the per-event signature is consistent with the empty-prediction-and-vocabulary-narrowing pattern just described. The cost model's worked example assumed CoverAndMerge preserves recall (\cref{ssec:cost}); on subscription distributions whose redundancy yields a CoverAndMerge representative set near or beyond the LLM's effective discrimination capacity, that assumption is empirically false.

The mechanism is consistent with the discrimination-capacity finding (\cref{ssec:discrim}): an LLM faced with subscription descriptions that each combine several distinct concepts must perform a finer-grained discrimination per matching decision than against unmerged subscriptions of the same total cardinality. The cost model accurately predicts \emph{invocation count} per cluster (\cref{fig:cost-validation}) but not \emph{accuracy} when CoverAndMerge produces compound descriptions. A refined model would condition the compression payoff on the redundancy structure of the subscription set: high-redundancy sets (where merging produces semantically clean representatives) should benefit; low-redundancy sets and adversarial duplication-with-rename distributions (where merging concatenates unrelated concepts) should fall back to A0. Identifying the redundancy threshold at which A4 ceases to dominate A0 is left to follow-on work.

\subsection{Cross-Dataset Comparison and Discrimination-Capacity Panel}\label{ssec:res-cross}

\begin{table}[htbp]
\centering
\caption{Best Neural Router configuration per backend vs.\ the two strongest baselines: macro-F1 across the three datasets, best config in parentheses. $^{\dagger}$Sonnet single-seed, indicative. Full metrics (FPR, latency, all seven baselines) in Suppl.\ Tab.~1. D1\,=\,CardiffNLP, D2\,=\,EUR-Lex, D3\,=\,CASAS.}
\label{tab:cross-dataset-main}
\small
\begin{tabular}{@{}lccc@{}}
\toprule
\textbf{Method} & \textbf{D1} & \textbf{D2} & \textbf{D3} \\
\midrule
NR / Claude Haiku             & 0.656 (A0) & 0.089 (A1) & 0.401 (A4) \\
NR / Qwen-2.5-7B              & 0.505 (A0) & 0.045 (A1) & 0.320 (A6) \\
NR / Claude Sonnet$^{\dagger}$ & 0.717 (A0) & 0.316 (A0) & --         \\
\midrule
Sentence-BERT cosine          & 0.423      & 0.154      & 0.225      \\
TF-IDF cosine                 & 0.102      & 0.162      & 0.268      \\
\bottomrule
\end{tabular}
\end{table}

\begin{table}[htbp]
\centering
\caption{D2 discrimination-capacity panel ($|\mathcal{S}|{=}201$, 5 seeds; A0 = raw LLM, A1 = clustering only), F1 as $\mathrm{mean}_{\pm \mathrm{half\text{-}CI95}}$ (A0 variance zero: temperature-0, seed-invariant). $^{\dagger}$Sonnet single-seed on a 5K subsample (indicative). Row blocks (top to bottom): Qwen-2.5 same-family sweep (1.5/7/32B), Mistral-7B at matched 7B scale, pairwise embedding baselines, closed-API frontier.}
\label{tab:d2-discrim}
\small
\setlength{\tabcolsep}{4pt}
\begin{adjustbox}{max width=\columnwidth}
\begin{tabular}{@{}lrrr@{}}
\toprule
\textbf{Backend} & \textbf{Params} & \textbf{A0 F1} & \textbf{A1 F1} \\
\midrule
Qwen 2.5  1.5B       & 1.5\,B & $0.000_{\pm 0.000}$          & $0.001_{\pm 0.001}$ \\
Qwen 2.5  7B         & 7\,B   & $0.002_{\pm 0.000}$          & $0.045_{\pm 0.016}$ \\
Qwen 2.5  32B        & 32\,B  & \textbf{$0.127_{\pm 0.000}$} & $0.116_{\pm 0.022}$ \\
Mistral 7B Instruct  & 7\,B   & $0.039_{\pm 0.000}$          & $0.080_{\pm 0.015}$ \\
\midrule
Sentence-BERT cosine & --     & 0.154          & 0.154 \\
TF-IDF cosine        & --     & 0.162          & 0.162 \\
Claude Sonnet$^{\dagger}$ & --     & 0.316          & --    \\
\bottomrule
\end{tabular}
\end{adjustbox}
\end{table}

\Cref{tab:cross-dataset-main} reports the best Neural Router configuration per backend and the two strongest baselines across the three datasets (full metrics, FPR, latency, and all seven baselines in Suppl.\ Tab.~1). On D1 ($|\mathcal{S}|{=}19$, low semantic gap) and D3 ($|\mathcal{S}|{=}19$, very high semantic gap requiring abductive reasoning), the best LLM configuration outperforms all baselines (\cref{tab:cross-dataset-main}). On D2 ($|\mathcal{S}|{=}201$, moderate semantic gap), the picture is more nuanced: small open-weight models collapse below pairwise baselines, while larger or newer-generation models partially escape (\cref{tab:d2-discrim}). The four-model open-weight panel on D2 quantifies this gradient.

The panel (\cref{tab:d2-discrim}) exhibits two structured patterns. First, \emph{same-family parameter scaling}: within Qwen 2.5, raw-LLM (A0) F1 rises monotonically across 1.5B$\to$7B$\to$32B (a $>$60$\times$ jump from 7B to 32B). Second, at \emph{matched parameter scale}, the more recently trained Mistral 7B reaches an order of magnitude above Qwen 2.5 7B. The pairwise embedding baselines sit \emph{above} all four open-weight points (Qwen 32B is closest but does not surpass them); only the closed-API Sonnet reference ($\dagger$, single-seed) is clearly above the pairwise band. All values in \cref{tab:d2-discrim}.

These results refine the discrimination-capacity boundary identified in \cref{ssec:cost}: $|\mathcal{S}|_{\mathrm{cross}}$ is not a global scalar but a model-dependent threshold determined by the joint of parameter count and training generation. On D2 specifically, $|\mathcal{S}|=201$ is well above $|\mathcal{S}|_{\mathrm{cross}}$ for Qwen 1.5B and 7B (deep collapse, F1 essentially zero); is partially binding for Mistral 7B and Qwen 32B (substantially lifted F1 relative to the smaller Qwen models, but still below the model-independent pairwise baselines); and is comfortably below $|\mathcal{S}|_{\mathrm{cross}}$ only for the closed-API frontier (Sonnet, the only point clearly above pairwise baselines in our panel). The capability-dependent gradient is consistent with the attention-budget mechanism but requires a second multiplicative factor --- instruction-tuning sophistication / training generation --- to account for the Mistral-vs-Qwen-7B gap at fixed parameter count.

\textbf{Statistical significance.}
A0 macro-F1 is a deterministic function of the matching prompt under temperature-0 decoding and is therefore identical across all five $k$-means seeds for every panel cell, so the variance reported in \cref{tab:d2-discrim} for A0 reflects subscription-set partitioning only and collapses to zero for unclustered runs. The Wilcoxon signed-rank test on the three headline pairs at A0 (Mistral 7B vs.\ Qwen 7B, Qwen 32B vs.\ Qwen 7B, Qwen 1.5B vs.\ Qwen 7B) returns the minimum-power $p{=}0.0625$ at $n{=}5$ paired observations: the inter-seed differences are all of identical sign at every cell, so the test cannot distinguish them from any reproducible directional effect. The two headline gaps (Mistral-over-Qwen at matched 7B scale, Qwen-32B-over-7B same-family) therefore rest on consistent directionality across all five seeds rather than on the formal $\alpha{=}0.05$ rejection. The A1 cells (where clustering randomness produces seed-to-seed variation) carry tighter Wilcoxon power once larger $n$ is collected; this is on the deferred-experiments list.

\subsection{QoE Heterogeneous Backend Assignment}\label{ssec:res-qoe}

Suppl.\ Tab.~2 compares the three backend assignment strategies defined in \cref{ssec:qoe-exp} on CardiffNLP with two open-weight Qwen-2.5 tiers (7B and 32B). All four observations are clear from the table:

\emph{(i) The bigger tier dominates on F1.} Homogeneous Qwen-2.5-32B reaches F1${=}0.225 \pm 0.019$, ahead of Qwen-2.5-7B at F1${=}0.166 \pm 0.023$, at the cost of $\sim 2.5{\times}$ longer wall-clock latency ($359$\,s vs.\ $141$\,s on a $1{,}000$-event subsample). \emph{(ii) Round-robin almost matches the bigger tier on F1 at $0.7{\times}$ its latency.} Round-robin (alternating clusters between Qwen-7B and Qwen-32B) yields F1${=}0.215 \pm 0.006$ at $L{=}250$\,s, closing $83\%$ of the $0.166{\to}0.225$ F1 gap while roughly halving the smaller-to-larger latency penalty. \emph{(iii) QoE-optimised differentiates assignments by weight preset, but does not outperform round-robin in this campaign.} The three weight presets produce three distinct per-cluster assignments (verified seed-by-seed; the bug-fixed implementation uses per-cluster min-max normalisation across candidate backends, so weights flip the argmax whenever F1 and cost or latency disagree on a cluster), but on D1 with a $10\%$ calibration sample, the resulting F1 ($0.179$--$0.192$) sits below both round-robin and the larger homogeneous tier. The cost-first preset attains the highest F1 among QoE variants ($0.192 \pm 0.039$) because it routes more clusters to the lower-token-count backend (Qwen-32B emits shorter responses than Qwen-7B in our prompt format), which happens to be both the cheaper and the more accurate option in this campaign. \emph{(iv) The bottleneck is calibration noise --- a hypothesis the sweep below tests and bounds.} Per-cluster F1 estimated from $10\%$ of events ($\sim 67$ events per cluster $\times$ 2 backends) carries enough variance that the greedy argmax frequently selects the worse backend at the cluster level; round-robin avoids this by skipping the calibration step entirely. The natural conjecture --- that QoE-optimised overtakes round-robin once a larger calibration sample drives per-cluster F1 standard error below the inter-backend gap --- is tested directly by the calibration-fraction sweep of \cref{ssec:res-qoe-calfrac}: across $\mathrm{frac}\in\{0.05,\ldots,1.00\}$ \emph{no} fraction crosses $\alpha{=}0.05$ (Friedman $p{=}0.290$), so the strong calibration-noise-limited claim is \emph{not} empirically supported on this 7B/32B gradient. Where the QoE framework does pay off is the cost-side localisation established under perturbation (\cref{ssec:res-qoe-pert}): separation from round-robin at the \texttt{cost\_first} preset only. On this gradient round-robin is a strong calibration-free baseline.

\subsubsection{Perturbation: workload shift and latency injection}\label{ssec:res-qoe-pert}

\begin{figure}[t]
  \centering
  \includegraphics[width=0.65\columnwidth]{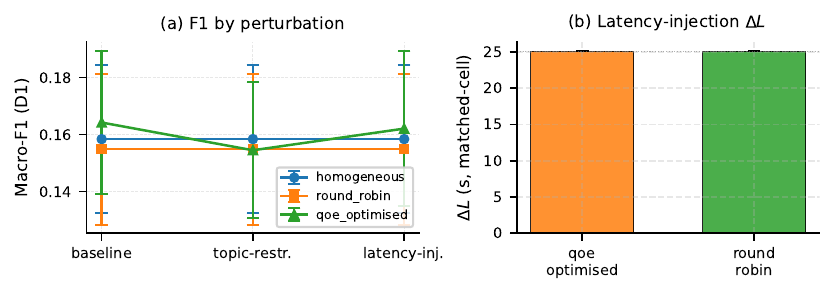}
  \caption{Perturbation results on D1, Qwen-2.5 7B/32B tier gradient, calibration fraction $0.10$, $n{=}15$ seeds. \emph{Left:} per-(perturbation, strategy) macro-F1 with 95\,\% CIs. \emph{Right:} matched-cell latency delta (latency-injection $-$ baseline) on mixed-backend strategies recovers the $0.05\,\mathrm{s}\times 500$\,event injection as ${+}25$\,s. Calibration-free strategies show zero matched-cell F1 delta on topic-restricted by construction.}
  \label{fig:qoe-perturbation}
\end{figure}

We exercise the QoE loop on D1 with two perturbations under the matched-pair regime defined in \cref{ssec:qoe-exp} ($n{=}15$ seeds, calibration fraction $0.10$; \cref{fig:qoe-perturbation}). \emph{Topic-restricted calibration} draws the calibration sample from three of nineteen CardiffNLP topics, simulating calibration-vs-deployment distribution shift; \emph{latency injection} adds a $0.05$\,s delay to every LLM call for events with index $\geq 500$. Matched-cell F1 deltas are exactly zero for the calibration-free strategies ($\Delta F1{=}0.000$ for both homogeneous and round-robin, as theoretically required), and QoE-optimised moves in the predicted direction with $\Delta F1{=}{-}0.018$ (95\,\% CI $[{-}0.031, {-}0.005]$, paired Wilcoxon $p{=}0.008$), consistent with the calibration-quality hypothesis. For latency injection, the matched-cell latency delta on the mixed-backend strategies is $\Delta L{=}{+}25.08$\,s (95\,\% CI $[25.06, 25.10]$, $n{=}14$; one seed dropped due to a missing baseline-vs-injection cell, paired Wilcoxon $p{<}10^{-3}$), exactly matching the expected $0.05\,\mathrm{s} \times 500$ post-injection events; matched-cell $\Delta F1$ on the same comparison is $0.000$ ($[{-}0.017, {+}0.018]$), confirming the perturbation cleanly isolates the latency dimension. Together the two perturbations close the ``exercise-the-mechanism'' gap. Stratified by weight preset, QoE-optimised separates from round-robin only at \texttt{cost\_first} (F1 weight $0.15$): paired $\Delta F1$(QoE${-}$RR)${=}{+}0.021$ on baseline (one-sided Wilcoxon $p{=}0.018$, $11/4$ positive seeds, $n{=}15$) and ${+}0.019$ under latency injection ($p{=}0.034$, $n{=}14$); broken by topic-restricted calibration ($p{=}0.28$). At \texttt{accuracy\_first} and \texttt{balanced}, QoE does not separate from round-robin --- the 7B/32B per-cluster F1 gap is too small for selective routing to outweigh calibration noise. Replacing the 32B tier with Qwen-2.5-72B (same $n{=}15$, frac${=}0.10$) widens the homog F1 gap from $0.029$ to $0.042$; the $\alpha{=}0.05$ separation does not generalise (\texttt{accuracy\_first} leads with $11{/}4$ positives, $p{=}0.126$), but QoE-cost-first matches homog-72B within $0.012$ F1 at $\Delta L{=}{-}738$\,s, preserving the tunable tradeoff (full perturbation companion in Suppl.\ §F).

\subsubsection{Calibration-fraction sweep}\label{ssec:res-qoe-calfrac}

We test the calibration-noise-limited claim by sweeping the calibration sample fraction $\in \{0.05, 0.10, 0.20, 0.50, 0.80, 1.00\}$ on D1 with the Qwen-2.5 7B / 32B tier gradient (Suppl.\ Fig.~2; matched-pair regime, $n{=}15$ seeds, balanced preset). Calibration-free strategies are flat and QoE-optimised's mean trajectory is non-monotone (F1 $0.150$--$0.165$); the paired one-sided Wilcoxon (QoE $>$ round-robin) crosses no fraction (six-fraction Friedman $\chi^2{=}6.17$, $p{=}0.290$; per-fraction $p$-values in Suppl.\ Fig.~2). \emph{No fraction crosses $\alpha{=}0.05$}: the calibration-noise-limited claim, in its strong form, is empirically unsupported on this tier gradient. The cost-side separations from \cref{ssec:res-qoe-pert} (QoE $>$ round-robin only at \texttt{cost\_first}) localise where the framework pays off.

The QoE framework therefore contributes a tunable, weight-driven per-cluster routing mechanism that correctly differentiates assignments by operator preference; round-robin is a strong calibration-free baseline competitive with the larger homogeneous tier at $0.7\times$ its latency. Across the perturbation (\cref{ssec:res-qoe-pert}) and calibration-fraction (\cref{ssec:res-qoe-calfrac}) experiments, QoE-over-round-robin separation is localised at the \texttt{cost\_first} preset and no calfrac crosses $\alpha{=}0.05$; the calibration-sample-to-discrimination-gap ratio is supported as the operative knob in its weak (sign-of-effect, cost-side-localised) form on this tier gradient.

\section{Discussion}\label{sec:discussion}

The ablation confirms the cost model's crossover prediction: at the evaluated volumes ($|\mathcal{S}|{\leq}201$) the simplest configuration A0 is the best Neural Router configuration (shifting to A1 on D2 for the open-weight panel/Haiku and A4 on D3 for Haiku; \cref{ssec:res-ablation,ssec:res-cross}). A0's dominance is predicted, not a pipeline failure: above the context-window crossover compression can only remove information, and modern cloud LLMs ($W{\geq}128$K) sit above it for $|\mathcal{S}|{\leq}201$; the pipeline's value is the sub-crossover regime (edge models, long $|\mathcal{S}|$), where the empirical $W{=}4{,}096$ sweep (\cref{ssec:crossover}) refines the picture --- at high $|\mathcal{S}|$ on duplication-with-rename sets A4 falls below A0 via empty-prediction collapse, so the recall-preservation assumption is conditional on the LLM engaging with the merged set. Two readings follow. \emph{Backend choice dominates configuration choice where $|\mathcal{S}|$ binds}: on the high-$|\mathcal{S}|$ D2 panel the cross-LLM F1 gap exceeds the within-LLM spread across non-dominated configurations (\cref{ssec:res-cross}); there, select the LLM first and tune the pipeline second. \emph{Dataset characteristics modulate the LLM advantage beyond what the cost model predicts}: D3 ($|\mathcal{S}|{=}19$) succeeds across all backends while D2 ($|\mathcal{S}|{=}201$) degrades model-dependently, so cardinality alone is not the whole story --- the binding factor is $|\mathcal{S}|$ relative to each model's discrimination capacity, which we analyse next.

\subsection{Discrimination Capacity}\label{ssec:discrim}

The two-crossover framework of \cref{ssec:cost} ($W_{\mathrm{cross}}$, $|\mathcal{S}|_{\mathrm{cross}}$) admits empirical localisation only for $|\mathcal{S}|_{\mathrm{cross}}$: $W_{\mathrm{cross}}$ is computed from the cost model, while $|\mathcal{S}|_{\mathrm{cross}}$ is model-dependent and must be characterised by sweep. The D2 panel (\cref{ssec:res-cross}, $|\mathcal{S}|{=}201$, four open-weight models) localises it along two axes --- same-family parameter scaling (Qwen 2.5 1.5B$\to$7B$\to$32B) and training generation at matched scale (Mistral 7B vs.\ Qwen 2.5 7B), per-model F1 in \cref{ssec:res-cross} --- with only the closed-API frontier crossing the pairwise-baseline band. D3 has the highest semantic gap yet succeeds at $|\mathcal{S}|{=}19$ across all backends, so semantic difficulty alone does not explain the D1$\to$D2 gap: the binding factor is $|\mathcal{S}|$ relative to the model's discrimination capacity.

The underlying mechanism --- a shared attention budget $\mathcal{O}(1/|\mathcal{S}|)$ scaled by parameter count, compounded by instruction-tuning generation (so Mistral $\gg$ Qwen-7B at matched scale), which pairwise methods avoid because $(e,s)$ comparisons are independent --- and the resulting deployment implications (model selection upstream of pipeline tuning; explicit $D$-management via subscription partitioning or embedding-prefilter cascades where no model fits below the boundary) are detailed in the electronic supplement. A quantitative scaling law for $|\mathcal{S}|_{\mathrm{cross}}$ is left to follow-on work.

\subsection{The Accuracy--Latency Trade-off}\label{ssec:tradeoff}

LLM-based matching is more expensive per invocation than syntactic or embedding-distance filters; the Neural Router accepts this for semantic understanding, with the mitigation derived formally in \cref{ssec:cost} (\cref{eq:latency}): \emph{cover/merge compression} reduces invocations below the context-window crossover, \emph{batching} amortises per-invocation overhead across $b_{\max}$ events, and \emph{parallelism} across $P$ instances converts the chain into $R{=}\lceil I/P\rceil$ rounds. The system targets regimes where per-batch latency on the order of seconds is acceptable (analytics dashboards, notification feeds, content recommendation, regulatory-alert dissemination, network-telemetry routing); rapid LLM-inference improvements~\cite{kwon2023vllm,zheng2024sglang} are narrowing the gap.

\subsection{Limitations}\label{ssec:limitations}

\textbf{Experimental scope.} The matching engine is evaluated as offline top-$\kappa$ multi-label retrieval on static labelled corpora, single-broker, single-tenant; events do not arrive as a stream, subscriptions do not churn, and there is no broker-level concurrency or backpressure. The seven baselines target the matching task itself; we do not include a traditional CBPS broker (Siena, JEDI, OpenPubSub) in the panel because their structured-filter languages are not designed to match the modality-gap setting D3 stresses. All inference runs on CSC's Mahti and Puhti A100/V100 partitions; the ``computing-continuum'' framing describes the design's intended deployment span, not the experimental hardware. Generalisation to online broker integration, churning subscription sets, and heterogeneous edge silicon is a deployment-study agenda left to follow-on work.

\textbf{Inference latency and cost.} Each LLM invocation adds latency and per-token cost; \cref{ssec:tradeoff} and the token-count cost model (\cref{eq:tokencost}) quantify the compression/batching/parallelism mitigation and let practitioners apply provider-specific pricing across the evaluated spectrum (zero-API-cost local Qwen/Mistral through Haiku/Sonnet).

\textbf{Cost-model fit on the non-trivial stratum.} The $83\%$ within-band headline (\cref{ssec:res-cost}) aggregates a strong fit on trivial cells ($m_c{\le}b_{\max}$; $48/81$, zero violations, model collapses to identity) with a weaker fit on the $33$ non-trivial cells where the ceiling $\lceil m_c/b_{\max}\rceil$ is genuinely under test ($14/33{=}42\%$ violations). A variance-aware revision (predicting the per-cluster ratio variance) is the natural extension; we report the stratified breakdown for full disclosure.

\textbf{Modality, language, and assignment scope.} The three datasets span social media, legal, and IoT smart home, but all use English text (sensor events are templated to text). Native multi-modal matching, non-English, and highly technical domains (medical, code) remain to be validated. The QoE comparison (Suppl.\ Tab.~2) is single-tenant, batch-evaluation, no-failure, so round-robin's load-balancing weaknesses (skew, heterogeneous throughput, failure, SLAs) are not exercised; the per-cluster min-max normalisation (Eq.~\ref{eq:qoe}) is candidate-set dependent. The QoE mechanism assumes per-cluster accuracy, cost, and latency are measured by the operator from held-out calibration data, \emph{not} self-reported by competing backend providers; the strategic regime in which a provider misreports to influence assignment is explicitly out of scope here and is a credibility-mechanism question (auditable logs, commitment devices) deferred to the service-economy treatment of~\selfcite{loven2026service}.

\textbf{Benchmark contamination.} D1 (CardiffNLP, 2022), D2 (EUR-Lex MultiEURLEX, 2021), and D3 (CASAS hh113, pre-2024) are all public datasets with release dates predating the training cut-offs of the LLMs evaluated here. Some fraction of the headline F1 numbers may therefore measure surface-level memorisation rather than zero-shot semantic matching. We do not run a closed-book probe in the current campaign; numbers in \cref{tab:ablation}, \cref{tab:d2-discrim}, and the cross-dataset table in the supplementary material should accordingly be read as upper bounds for genuinely novel deployments. A held-out post-cut-off slice (e.g., post-2024 EU legislation with EUROVOC tags, or a synthetic CASAS template set on a held-out sensor schema) would tighten the external-validity argument and is on the deferred-experiments list.

\subsection{Broader Implications and Future Work}

The Neural Router shows LLMs can serve as practical matching engines for content-based pub/sub, not only as downstream consumers. As autonomous agents compose, negotiate, and consume services across device-edge-cloud environments~\cite{saleh2025towards}, they cannot control the vocabulary or format of those services; the matching engine bridges vocabulary gaps via LLM-driven reasoning rather than pre-engineered taxonomies, with D3 instantiating the modality-gap regime where embedding baselines fail by construction. The most consequential extensions cluster around three themes: \emph{(i)}~a quantitative scaling law for $|\mathcal{S}|_{\mathrm{cross}}$ in parameters~$\times$~training generation, validated across cross-family open-weight points (Gemma, Phi, future Llama), with discrimination-aware partitioning ($|\mathcal{S}'|\leq D$ per prompt) and embedding-prefilter cascades for high-$|\mathcal{S}|$ regimes; \emph{(ii)}~online operation --- incremental subscriptions, multi-modal extensions, distributed federation across administrative domains, and operational robustness against silent-snapshot rotation; \emph{(iii)}~convergence with service-economy mechanisms when the engine is offered as a priced service~\selfcite{loven2026service} --- formal capacity allocation, credibility mechanisms (auditable logs, commitment devices), and quality-of-experience menus that elicit private valuations. Together these extend the Neural Router from a single-broker prototype toward a deployable component of decentralised agentic service ecosystems.

\section{Conclusion}\label{sec:conclusion}

We introduced the Neural Router, a content-based publish/subscribe matching engine integrating LLMs into the matching loop, characterised by a two-crossover cost-accuracy model (context-window and discrimination-capacity), three composable algorithms (\textsc{OptimizeSubscriptions}, \textsc{MatchEvents}, \textsc{CoverAndMerge}), and a QoE-based heterogeneous backend assignment mechanism. Evaluation across three datasets (CardiffNLP, EUR-Lex, CASAS), six LLM backends (Qwen 2.5 1.5B/7B/32B, Mistral 7B, Haiku, Sonnet), and seven ablations against seven baselines confirms the cost model: raw LLM matching (A0) dominates above the context-window crossover (D1 Haiku F1\,=\,0.66); below it, a $W{=}4$K sweep shows compression collapsing 0.37$\to$0.04 as $|\mathcal{S}|$ grows 50$\to$2{,}000 via empty-prediction collapse and vocabulary narrowing. The invocation-count prediction holds per-cluster on $n{=}81$ cells (median $1.00$, $83\%$ in band; misses concentrate in $33$ non-trivial cells). The D2 four-model open-weight panel reveals a structured discrimination-capacity escape consistent with a joint parameter-count $\times$ training-generation factor: Qwen 1.5B/7B collapse, Mistral 7B and Qwen 32B partially escape, only the closed-API frontier crosses the pairwise embedding band. The two-crossover model gives practitioners a principled cost-side tool plus an empirically characterised viability boundary; deployment validation, multi-modal extensions, distributed federation, market-mechanism convergence, and a quantitative $|\mathcal{S}|_{\mathrm{cross}}$ scaling law are the natural follow-ons.

\begin{acks}
This work was supported by Business Finland through the Neural Pub/Sub project (Diary No.\ 8754/31/2022), by the Research Council of Finland (Grant No.\ 362594), by the 6G Flagship program (Grant No.\ 369116), and by the Digital Twinning of Personal Area Networks for Optimized Sensing and Communication project (Diary No.\ 8782/31/2022).
\end{acks}

\section*{Data and Software Availability}
The datasets used in this study (CardiffNLP Tweet Topic, EUR-Lex MultiEURLEX, CASAS hh113) are publicly available from their respective sources cited in \cref{ssec:dataset}. The Neural Router source code, DVC pipeline definitions, experiment configurations, regression tests, and per-event prediction parquet files are available at \url{https://github.com/lloven/neural-router-experiments}. The repository is under access embargo during peer review; reviewers may obtain read access via the Guest Editors on request. Upon acceptance the repository will be made public and a Zenodo DOI minted from the frozen snapshot recorded in the artefact's \texttt{paper.tag} file, replacing this URL at camera-ready.

\bibliographystyle{ACM-Reference-Format}
\bibliography{bibtex/neural,bibtex/new-ref}

\appendix

\section{Prompt templates}\label{app:prompts}

The Neural Router uses two prompt templates referenced from the body's Prompt Design subsection (\S 3.4 in the body): \cref{fig:listing1} is the subscription-optimisation (cover/merge) prompt consumed by the \textsc{CoverAndMerge} algorithm (Algorithm~4 in the body); \cref{fig:listing2} is the event-matching prompt that implements the matching function $\mu$ from the body's Problem Statement (\S 3.1) and is invoked by the \textsc{MatchEvents} algorithm (Algorithm~3 in the body). Both use Python f-string placeholders for runtime injection of cluster subscriptions and event batches.

\begin{figure}[t]
\begin{lstlisting}[style=prompt,caption={Subscription-optimisation (cover/merge) prompt template. \texttt{\{cluster\_subs\}} is replaced with the cluster's $|c.\mathcal{S}'|$ subscriptions formatted one per line as ``\texttt{[id] description}''. The LLM returns JSON \texttt{\{"covers": [[i,j], ...], "merges": [[i,j], ...]\}}. Decoding: temperature 0, max\_tokens 512. See \textsc{CoverAndMerge} (Algorithm~4 in the body).},label={fig:listing1}]
SYSTEM: You are a content-router optimiser. Given a list of natural-language
subscription descriptions, identify pairs (i, j) where subscription i covers
j (i.e., every event matching j also matches i) and pairs (i, j) where two
subscriptions overlap enough to be merged into a single combined description.
Return JSON only.

USER:
Subscriptions:
{cluster_subs}

Return JSON with keys "covers" and "merges", each a list of [i, j] index
pairs. Use indices from the list above (1-based). Return {} if no pair
qualifies.
\end{lstlisting}
\begin{lstlisting}[style=prompt,caption={Event-matching prompt template. \texttt{\{active\_subs\}} is the post-CoverAndMerge subscription set for the cluster (or the raw set, for A0); \texttt{\{event\_batch\}} is the current batch of $b \le b_{\max}$ events. The LLM returns JSON \texttt{\{"matches": [[event\_idx, sub\_id], ...]\}}, with up to $\kappa{=}3$ subscriptions per event. Decoding: temperature 0, max\_tokens scaled by batch size. See \textsc{MatchEvents} (Algorithm~3 in the body).},label={fig:listing2}]
SYSTEM: You are a content router. For each event below, return up to K
subscription IDs whose descriptions match the event's content. Return JSON
only; no commentary.

USER:
Subscriptions ([id] description):
{active_subs}

Events:
{event_batch}

Return JSON of the form {"matches": [[event_idx, sub_id], ...]} where
event_idx is 1-based and sub_id is the [id] from the subscription list.
At most K=3 subscriptions per event; omit events with no matching
subscription.
\end{lstlisting}
\end{figure}

\section{Parameter-sensitivity panel}\label{app:sensitivity}

\begin{figure}[t]
  \centering
  \begin{subfigure}[t]{0.48\linewidth}
    \centering
    \includegraphics[width=\linewidth]{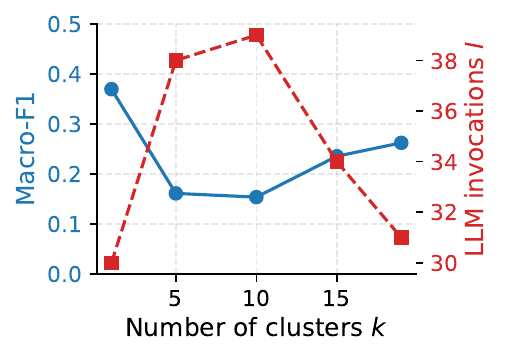}
    \caption{Cluster count $k$.}
    \label{fig:k-sensitivity}
  \end{subfigure}\hfill
  \begin{subfigure}[t]{0.48\linewidth}
    \centering
    \includegraphics[width=\linewidth]{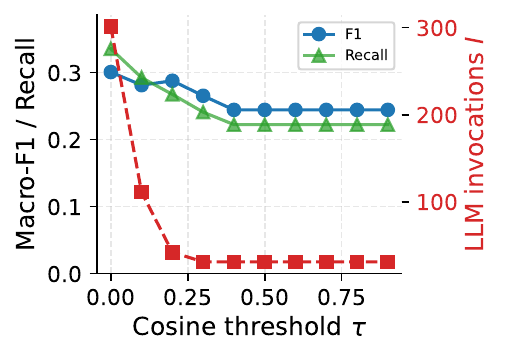}
    \caption{Cosine threshold $\tau$.}
    \label{fig:tau-sensitivity}
  \end{subfigure}

  \vspace{4pt}

  \begin{subfigure}[t]{0.48\linewidth}
    \centering
    \includegraphics[width=\linewidth]{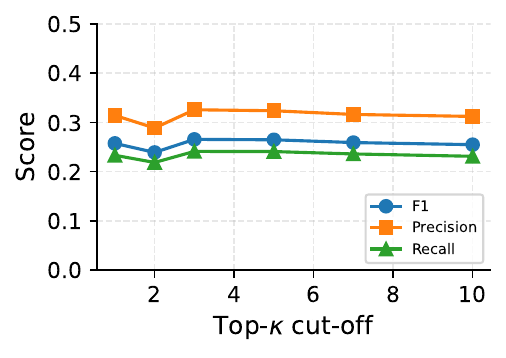}
    \caption{Top-$\kappa$ cut-off.}
    \label{fig:kappa-sensitivity}
  \end{subfigure}\hfill
  \begin{subfigure}[t]{0.48\linewidth}
    \centering
    \includegraphics[width=\linewidth]{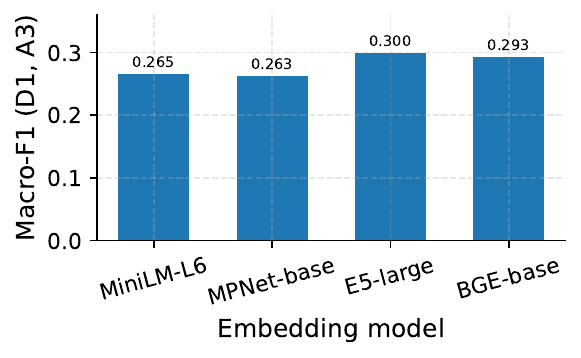}
    \caption{Embedding model choice.}
    \label{fig:embedding-sensitivity}
  \end{subfigure}
  \caption{Parameter sensitivity on D1 (Haiku, A4): F1 vs.\ each parameter holding the others at the production defaults from §4.2 in the body. The plateau in $\tau$ above $\approx 0.3$ motivates the discussion of cosine-prefilter geometry in §6.1 in the body.}
  \label{fig:sensitivity-panel}
\end{figure}

\Cref{fig:sensitivity-panel} accompanies the parameter-sensitivity discussion in §5.2 in the body. The four panels sweep, respectively, the cluster count $k$, the cosine pre-filter threshold $\tau$, the top-$\kappa$ cut-off, and the embedding-model choice on D1 with the Haiku backend in configuration A4.

\section{Cross-dataset comparison table}\label{app:cross-dataset}

\Cref{tab:cross-dataset} reports the best Neural Router configuration per backend against all seven baselines across the three datasets, complementing the dataset-specific narratives in §5.5 in the body and §6 in the body.

\begin{table*}[t]
\centering
\caption{Cross-dataset comparison: best Neural Router configuration per backend vs.\ baselines. F1 reported as $\mathrm{mean}_{\pm \mathrm{half\text{-}CI95}}$ over 5 seeds with the winning config in parentheses; $^{\dagger}$Sonnet rows are single-seed and indicative only. D1\,=\,CardiffNLP (19 topics, short text); D2\,=\,EUR-Lex (201 subjects, long legal); D3\,=\,CASAS smart home (19 activities, IoT sensor). Qwen-2.5-7B runs use the stratified-subsample event caps documented in §4.2 in the body; Haiku and pairwise baselines use the full corpus; Sonnet D2 uses the 5{,}000-event subsample. DistilBART-MNLI uses the full D1 corpus and stratified subsamples on D2 ($|\mathcal{E}|{=}200$) and D3 ($|\mathcal{E}|{=}1{,}000$). Sonnet D3 was not evaluated; see §6.3 in the body.}
\label{tab:cross-dataset}
\small
\setlength{\tabcolsep}{3pt}
\begin{adjustbox}{max width=\textwidth}
\begin{tabular}{@{}l ccc ccc ccc@{}}
\toprule
& \multicolumn{3}{c}{\textbf{F1 (best config)}} & \multicolumn{3}{c}{\textbf{FPR}} & \multicolumn{3}{c}{$\boldsymbol{L}$\textbf{\,(s)}} \\
\cmidrule(lr){2-4} \cmidrule(lr){5-7} \cmidrule(lr){8-10}
\textbf{Method} & \textbf{D1} & \textbf{D2} & \textbf{D3} & \textbf{D1} & \textbf{D2} & \textbf{D3} & \textbf{D1} & \textbf{D2} & \textbf{D3} \\
\midrule
NR / Claude Haiku       & $0.656_{\pm 0.001}$\,(A0) & $0.089_{\pm 0.020}$\,(A1) & $0.401_{\pm 0.011}$\,(A4) & 0.030 & 0.019 & 0.076 &  3.1 & 11.4 &  1.7 \\
NR / Qwen-2.5-7B        & $0.505_{\pm 0.000}$\,(A0) & $0.045_{\pm 0.016}$\,(A1) & $0.320_{\pm 0.001}$\,(A6) & 0.024 & 0.010 & 0.060 &  1.9 & 12.0 & 43.6 \\
NR / Claude Sonnet$^{\dagger}$ & 0.717\,(A0) & 0.316\,(A0) & --          & 0.038 & 0.015 & --    &  5.3 & 14.7 & --   \\
\midrule
BM25                       & 0.082 & 0.009 & 0.130 & 0.161 & 0.169 & 0.155 &      1.8 &     4.5 &     3.2 \\
Sentence-BERT cosine       & 0.423 & 0.154 & 0.225 & 0.115 & 0.131 & 0.144 &     21.5 &    35.7 &    26.9 \\
Cross-encoder              & 0.323 & 0.094 & 0.176 & 0.128 & 0.149 & 0.151 &    433.1 &      -- &      -- \\
DistilBART-MNLI zero-shot  & 0.434 & 0.077 & 0.094 & 0.114 & 0.022 & 0.156 & 10{,}323 &     678 &     413 \\
GloVe cosine               & 0.104 & 0.027 & 0.079 & 0.158 & 0.166 & 0.158 &      0.1 &     0.4 &     0.3 \\
TF-IDF cosine              & 0.102 & 0.162 & 0.268 & 0.158 & 0.130 & 0.138 &      0.2 &     0.6 &     0.5 \\
Word2Vec cosine            & 0.110 & 0.028 & 0.077 & 0.157 & 0.166 & 0.158 &      0.1 &     0.5 &     0.3 \\
\bottomrule
\end{tabular}
\end{adjustbox}
\end{table*}

\section{QoE backend-assignment table}\label{app:qoe}

\Cref{tab:qoe} reports the per-strategy F1, cost, and latency comparison summarised qualitatively in §5.6 in the body.

\begin{table}[t]
\centering
\caption{Heterogeneous backend assignment strategies on CardiffNLP, Qwen-2.5 tier sweep (tier\_mid$=$Qwen-2.5-7B, tier\_large$=$Qwen-2.5-32B). All rows use the A3 configuration on a 1{,}000-event stratified subsample, mean over 5 seeds. Cost is the GPT-4o-mini-equivalent token cost per 1{,}000 events; for self-hosted Qwen tiers the actual deployment cost is more accurately approximated by latency $L$. The QoE-optimised rows use per-cluster min-max normalisation across candidate backends (§4.9 in the body).}
\label{tab:qoe}
\small
\setlength{\tabcolsep}{4pt}
\begin{tabular}{@{}lccc@{}}
\toprule
\textbf{Strategy} & \textbf{F1 (mean)} & \textbf{\$/1k evt} & $\boldsymbol{L}$\textbf{\,(s)} \\
\midrule
\multicolumn{4}{@{}l}{\textit{Single-backend baselines}} \\
Homogeneous (Qwen-2.5-7B)  & $0.166_{\pm 0.023}$ & $0.017_{\pm 0.000}$ & $141_{\pm 5}$ \\
Homogeneous (Qwen-2.5-32B) & $\mathbf{0.225}_{\pm 0.019}$ & $0.015_{\pm 0.001}$ & $359_{\pm 84}$ \\
\midrule
\multicolumn{4}{@{}l}{\textit{Mixed-strategy assignment}} \\
Round-robin                 & $0.215_{\pm 0.006}$ & $0.016_{\pm 0.001}$ & $\mathbf{250}_{\pm 34}$ \\
QoE accuracy-first          & $0.179_{\pm 0.017}$ & $0.015_{\pm 0.002}$ & $246_{\pm 98}$ \\
QoE balanced                & $0.178_{\pm 0.032}$ & $0.015_{\pm 0.001}$ & $232_{\pm 31}$ \\
QoE cost-first              & $0.192_{\pm 0.039}$ & $0.015_{\pm 0.000}$ & $263_{\pm 37}$ \\
\bottomrule
\end{tabular}
\end{table}


\section{Calibration-fraction sweep figure}\label{app:qoe-calfrac}

\begin{figure}[t]
  \centering
  \includegraphics[width=0.6\columnwidth]{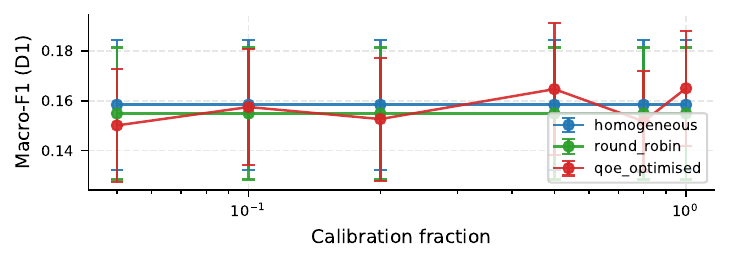}
  \caption{Calibration-fraction sweep on D1, Qwen-2.5 7B / 32B tier gradient, matched-pair LLM cache, $n{=}15$ seeds throughout. Paired one-sided Wilcoxon (QoE $>$ round-robin) returns $p{=}0.555, 0.381, 0.640, 0.180, 0.738, 0.126$ at frac${=}0.05$--$1.00$; six-fraction Friedman $\chi^2{=}6.17$, $p{=}0.290$. No fraction crosses $\alpha{=}0.05$: the calibration-noise-limited claim does not hold in its strong form on this tier gradient. Full numerical detail in body §5.6.2.}
  \label{fig:qoe-calfrac}
\end{figure}

\section{Wider-tier-gradient perturbation}\label{app:tier-72b-pert}

The body §5.6.1 reports a one-sentence summary of the Qwen-2.5 7B / 72B perturbation companion. We give the full table here so the body can be read without it.

The campaign mirrors the 7B/32B perturbation design (\texttt{baseline}, \texttt{topic\_restricted\_cal}, \texttt{latency\_injection}; calibration fraction $0.10$; $n{=}15$ seeds; matched-pair LLM cache; balanced QoE) and replaces the 32B tier with Qwen-2.5-72B Q4 served across two A100 40\,GB GPUs. The homogeneous F1 gap widens from $0.029$ (7B/32B) to $0.042$ (7B/72B; homog-mid $0.144_{\pm 0.039}$, homog-large $0.186_{\pm 0.064}$).

\textbf{Topic-restricted calibration.} Matched-cell $\Delta F1$(QoE-balanced)${=}{-}0.008$ ($n{=}15$, 95\,\% CI $[{-}0.024, {+}0.008]$, paired Wilcoxon $p{=}0.246$). The calibration-quality response detected at 7B/32B ($\Delta F1{=}{-}0.018$, $p{=}0.008$ in the body) is blunted at 7B/72B: the wider per-cluster F1 gap leaves the QoE argmax robust to a topic-biased calibration sample. The mechanism still operates --- the sign is still negative --- but the magnitude is no longer significant against the inflated inter-seed variance at the wider tier gradient.

\textbf{Latency injection.} The matched-cell latency delta is $\Delta L{=}{+}25.08$\,s as expected ($25.08{\pm}0.03$\,s on round-robin and QoE-optimised, $n{=}15$). Matched-cell $\Delta F1$(QoE-balanced)${=}{+}0.034$ ($n{=}15$, $p{=}0.0001$), unlike the clean isolation observed at 7B/32B ($\Delta F1{=}0.000$). This is not a measurement artefact: QoE's three-component objective uses latency, and during calibration the injected $0.05$\,s per event $\geq 500$ flows into the latency component of the per-cluster scalarisation. With min-max normalisation across two backends, the latency change is too small to flip the argmax at the 7B/32B latency ratio (12 vs 41\,s baseline) but does flip borderline clusters at the 7B/72B ratio (12 vs 49\,s). At this tier gradient, those flips happen to favour the better backend, so F1 rises with the injection. The reading: the ``clean dimension isolation'' of latency injection is contingent on the perturbation being below the resolution at which calibration argmax is sensitive, not a structural property of the QoE framework.

\textbf{Cost-and-latency tunability.} QoE-cost-first matches homogeneous-72B's F1 within $0.012$ at $\Delta L{=}{-}738$\,s relative to homogeneous-72B (matched-cell). The operator-tunable tradeoff is preserved across the wider gradient: QoE buys back almost all of the F1 ceiling at a fraction of the latency budget when the larger tier is genuinely costly to serve.

\section{Reproducibility apparatus: regression invariants}\label{app:reproducibility}

Five regression invariants are encoded as Python tests in the artefact repository (\texttt{tests/test\_figure\_data\_consistency.py}, \texttt{tests/test\_synthetic\_data\_metric\_invariant.py}, \texttt{tests/test\_calibration\_split\_invariant.py}) and run on every figure-rendering pass. They constrain the relationship between the underlying CSVs and the manuscript figures, so that a quiet data-side change cannot silently mis-align with the body claims.

\textbf{I1 (cost-model crossover monotonicity --- the $W{=}4{,}096$ context-budget crossover-validation diagnostic).} For $W{=}4{,}096$, A0 macro-F1 is non-decreasing in $|\mathcal{S}|$ until truncation engages, and constant thereafter. This is the operational signature of the cost-model crossover: below the truncation regime the prompt fits the budget and accuracy can only stay flat or improve as more subscriptions are seen; above it the prompt is truncated to the same prefix and accuracy plateaus.

\textbf{I2 (empty-prediction collapse --- the duplication-with-rename discrimination diagnostic).} The A4 empty-prediction rate is non-decreasing in $|\mathcal{S}|$ on duplication-with-rename subscription sets. This pins the discrimination-capacity diagnostic: as $|\mathcal{S}|$ grows under a fixed semantic complexity, the model increasingly emits empty predictions rather than a wrong answer.

\textbf{I3 (cost-model factor-of-two band --- the predicted-vs-measured invocation-count validation).} The cost-model factor-of-two band $|I_{\mathrm{pred}}/I_{\mathrm{meas}}-1| \le 1$ holds for at least $80\%$ of per-cluster cells. This formalises the predictive accuracy claim quoted in the body and breaks the build if a future re-run drifts the per-cluster distribution outside the headline tolerance.

\textbf{I4 (description-aware metric round-trip).} The description-aware F1 reduces to ID-based F1 when every ID has a distinct description. This is the metric-invariant test: the description-aware refinement only adds resolving power when descriptions disambiguate IDs; in the trivially distinct case it must reproduce the ID-based F1 exactly.

\textbf{I5 (calibration--evaluation disjointness).} For every (cluster, backend) cell in the QoE perturbation and calibration-fraction experiments, the calibration and evaluation event sets are disjoint, $E_{\mathrm{cal}}\cap E_{\mathrm{eval}}=\emptyset$, with the single intentional exception of the $\mathrm{frac}{=}1.00$ asymptotic-ceiling point (flagged in the body as an upper bound, not a generalisation estimate). This invariant guards the §5.6 perturbation and calibration-fraction conclusions against silent calibration--evaluation leakage at large calibration fractions.

The five invariants are TAAS-RCR-aligned: they are mechanically checkable, archived alongside the data and code, and their failure modes correspond directly to claims a reviewer might want to verify quickly.

\section{Discrimination-capacity mechanism and deployment implications}\label{app:discrim-detail}

This expands the discrimination-capacity finding of the body's Discrimination Capacity subsection; the headline localisation (binding factor $|\mathcal{S}|$ relative to model discrimination capacity, two-axis D2 panel) remains in the body.

\textbf{Mechanism.} Two coupled factors. \emph{Attention-budget}: $|\mathcal{S}|$ simultaneous decisions over a shared attention mechanism gives $\mathcal{O}(1/|\mathcal{S}|)$ per-subscription budget, analogous to ``lost in the middle''~\cite{liu2024lost}; parameter count enlarges the absolute pool. \emph{Instruction-following}: newer instruction-tuning regimes yield tighter structured-output behaviour, accounting for Mistral $\gg$ Qwen 7B at matched scale. Pairwise methods avoid both because $(e,s)$ comparisons are independent. The $\tau$ sensitivity (Suppl.\ Fig.~1(b)) shows the cosine pre-filter operating in a near-binary regime ($\tau \in [0.05, 0.2]$), so it cannot selectively route at high $|\mathcal{S}|$ either.

\textbf{Implications.} Three takeaways: \emph{(i)}~model selection is upstream of pipeline tuning at $|\mathcal{S}|\sim 200$ (the 32B-over-1.5B and Mistral-over-Qwen gaps exceed any algorithmic-configuration gap on D2); \emph{(ii)}~parameter count and training generation contribute roughly independently within a 20$\times$ parameter range; \emph{(iii)}~where no model fits below the discrimination boundary, the architecture must manage $D$ explicitly via subscription partitioning ($\leq D$ candidates per prompt) or embedding-prefilter cascades.


\end{document}